\documentclass[twocolumn,english,aps,prb,floats]{revtex4-2}
\usepackage[T1]{fontenc}
\usepackage[latin9]{inputenc}
\usepackage{babel}
\usepackage{amsmath}
\usepackage{amssymb}
\usepackage{graphicx}
\usepackage{esint}
\usepackage[unicode=true,
 bookmarks=false,
 breaklinks=false,pdfborder={0 0 1},backref=section,colorlinks=false]
 {hyperref}

\makeatletter

\newcommand{\lyxdot}{.}

\usepackage{babel}

\usepackage{bm}
\usepackage{wasysym}
\usepackage{breakurl}

\@ifundefined{textcolor}{}{%
 \definecolor{BLACK}{gray}{0}
 \definecolor{WHITE}{gray}{1}
 \definecolor{RED}{rgb}{1,0,0}
 \definecolor{GREEN}{rgb}{0,1,0}
 \definecolor{BLUE}{rgb}{0,0,1}
 \definecolor{CYAN}{cmyk}{1,0,0,0}
 \definecolor{MAGENTA}{cmyk}{0,1,0,0}
 \definecolor{YELLOW}{cmyk}{0,0,1,0}
}

\makeatother

\begin{document}
\title{Localized Spin-Wave Modes and Microwave Absorption in Random-Anisotropy
Ferromagnets}
\author{Dmitry A. Garanin and Eugene M. Chudnovsky}
\affiliation{Physics Department, Herbert H. Lehman College and Graduate School,
The City University of New York, 250 Bedford Park Boulevard West,
Bronx, New York 10468-1589, USA }
\date{\today}
\begin{abstract}
The theory of localized spin-wave excitations in random-anisotropy
magnets has been developed. Starting with a pure Heisenberg ferromagnet,
we study the evolution of standing spin waves in a finite-size sample
towards localized modes on increasing the strength of random anisotropy.
Profiles of the localized modes and their phases are analyzed and
visualized in a 2D sample. Localization length is obtained by several
methods and its dependence on random anisotropy is computed. The connection
between the localization of spin excitations and the broadband nature
of the absorption of microwave power by random-anisotropy magnets
is elucidated. 
\end{abstract}
\maketitle

\section{Introduction}

Ferromagnetic exchange, magnetic anisotropy, and dipole-dipole forces
are leading interactions that determine the properties of ferromagnets
at different spatial scales. At the shortest scales, the exchange,
which is usually the strongest interaction, aligns the neighboring
spins in one direction. That direction is determined by the magnetic
anisotropy due to the symmetry of the underlying crystal lattice.
At the largest scale, the dipole-dipole interaction breaks the system
into magnetic domains. A less trivial situation occurs in amorphous
and sintered ferromagnets where the directions of local magnetic anisotropy
axes are random. They exhibit glassy properties that have been intensively
studied over the last fifty years, see, e.g., Ref.\ \onlinecite{garchu-JPhys2022}
and references therein. Like for spin glasses, the ground state of
random-anisotropy (RA) magnets remains the subject of discussions
and controversies. Since seminal works of Larkin \citep{Larkin} and
Imry and Ma \citep{IM} it has been clear, however, that random pushes
of the magnetization by the RA must lead to a significant disordering
on a scale that is inversely proportional to some power of the RA
strength.

Somewhat less attention has been paid to spin excitations in the RA
magnets. In conventional ferromagnets with uniform magnetization,
the ac magnetic field can generate uniform ferromagnetic resonance
(FMR) and spin waves with a finite wavelength. When the RA magnet
is placed in a strong polarizing magnetic field, similar phenomena
can occur \citep{Saslow2018}, although the effect of disorder shows
in the FMR width and the damping of spin waves. The least understood
case, which is of the greatest interest for applications, is the one
without the polarizing field when the magnet is fully disordered.

Anderson localization of spin waves due to random anisotropy \citep{Bruinsma1986,Serota1988}
and, more generally, localization of Bose excitations in disordered
systems \citep{Ma-PRB1986,Zhang-PRB1993,Alvarez-PRL2013,Yu-AnnPhys2013}
has been studied in the past. The effect of randomly distributed lattice
defects on the localization of spin waves and magnonic transport has
been addressed in Ref.\ \onlinecite{Nowak2015}. Experimental evidence
of spin-wave localization was reported in amorphous \citep{Amaral-1993,Suran1-1997,Suran2-1997,Suran-1998}
and inhomogeneous \citep{McMichael-PRL2003} magnetic films, as well
as in heterostructures \citep{Loubens-PRL2007} and in films with
the inhomogeneous magnetic field created by a force microscope \citep{Du-PRB2014}.

The RA model assumes (see, e.g., Refs.\ \onlinecite{CSS-1986,CT-book,PCG-2015,garchu-JPhys2022}
and references therein) that spins interact via ferromagnetic exchange
but that directions of local magnetic anisotropy axes are randomly
distributed from one spin to another. When no material anisotropy
is introduced by the manufacturing process, an amorphous or sintered
magnet has no preference for the orientation of its magnetic moment.
Weak local RA cannot break the local ferromagnetic order but it makes
the vector of the magnetization wander around the magnet. This model
was successfully applied to the description of static properties of
amorphous magnets, such as the ferromagnetic correlation length, zero-field
susceptibility, the approach to saturation, etc. \citep{RA-book,Marin-MMM2020}.
It was shown that the presence of topological defects makes properties
of the model depend on the dimensionality of space: $d=1$ (thin wires),
$d=2$ (thin films), $d=3$ (bulk systems), and on the number of spin
components (XY model vs Heisenberg) \citep{PGC-PRL2014}.

In recent years, significant attention, driven by applications, has
been paid to the absorption of microwaves by nanocomposites made of
magnetic particles of various shapes and sizes, see, e.g., Ref.\ \onlinecite{Zeng-2020}
and references therein. Amorphous and sintered ferromagnets represent
the ultimate limit of densely packed nanoparticles with distributed
properties. In a recent paper \citep{GC-PRB2021} we demonstrated
that they provide strong broadband absorption of microwave radiation.
Evidence has been obtained that it occurs via the absorption of the
energy of the ac magnetic field by localized spin excitations.

In this article, we present a detailed study of such excitations and
their effect on microwave power absorption using advantages offered
by modern computer power that had not existed when the localization
of spin waves in RA ferromagnets was studied before. Instead of pre-selecting
the mechanism of Anderson localization via phase-coherent scattering
of spin waves \citep{Akkermans-2007}, we begin with exciting standing
spin waves in a 2D sample containing $10^{4}$ -- $10^{6}$ spins,
and then compute their evolution at finite temperature towards localized
oscillations on increasing the strength of the RA. Since the effects
we are considering occur on a spatial scale that is typically small
compared with the scale dominated by magnetic dipolar forces, interactions
included in our model are ferromagnetic exchange, random magnetic
anisotropy, and Zeeman interaction of spins with the ac magnetic field.
Our choice of the 2D system is determined by applications of thin
films as microwave absorbers.

The paper is organized as follows. The model, the numerical method,
the description of excitation modes, and the formulas for the power
absorption are introduced in Section II. Equations describing the
time evolution of the excitation modes are derived in Section III.
Numerical results on visualization of spin excitations in the RA magnet,
on the localization transition, phases of the localized modes, are
presented in Section IV. A separate edge pumping experiment is presented
in Sec. V. Brief review of the results and remarks on their relevance
to experiments are given in Section VI.

\section{The model, numerical method, excitation modes }

We consider a model of classical spin vectors $\mathbf{s}_{i}$ ($\left|\mathbf{s}_{i}\right|=1$)
on a square lattice
\begin{equation}
\mathcal{H}=-\frac{1}{2}\sum_{ij}J_{ij}\mathbf{s}_{i}\cdot\mathbf{s}_{j}-\frac{D_{R}}{2}\sum_{i}\left(\mathbf{n}_{i}\cdot\mathbf{s}_{i}\right)^{2}-\mathbf{H}\cdot\sum_{i}\mathbf{s}_{i},
\end{equation}
where $J_{ij}$ is the nearest-neighbors exchange interaction with
the coupling constant $J>0$, $D_{R}$ is the strength of the random
anisotropy, $\mathbf{n}_{i}$ are randomly directed unit vectors,
and $\mathbf{H}$ is the magnetic field. The dynamics of classical
spins is governed by the Landau-Lifshitz (LL) equation
\begin{equation}
\hbar\dot{\mathbf{s}}_{i}=\mathbf{s}_{i}\times{\bf H}_{{\rm eff},i}-\alpha{\bf s}_{i}\times\left(\mathbf{s}_{i}\times{\bf H}_{{\rm eff},i}\right),\quad{\bf H}_{{\rm eff},i}\equiv-\frac{\partial\mathcal{H}}{\partial\mathbf{s}_{i}}\label{Larmor}
\end{equation}
where $\alpha\ll1$ is the phenomenological damping constant due to
interaction of spins with electrons, phonons, etc. In the subsequent
computations, we set $\alpha=0$ as in the presence of RA and at nonzero
temperatures there is an intrinsic damping due to mode hybridization
and nonlinearity in the system. 

Numerical investigation of the localization of modes is performed
in several steps. First, a local energy minimum is found by the energy
minimization starting from a collinear or random state of the spins
-- collinear initial condition (CIC) or random initial condition
(RIC). The numerical method \citep{garchupro13prb} combines sequential
alignment of spins ${\bf s}_{i}$ with the direction of the local
effective field, ${\bf H}_{{\rm eff},i}$, with the probability $\eta$,
and the energy-conserving spin flips (overrelaxation), ${\bf s}_{i}\to2({\bf s}_{i}\cdot{\bf H}_{{\rm eff},i}){\bf H}_{{\rm eff},i}/H_{{\rm eff},i}^{2}-{\bf s}_{i}$,
with the probability $1-\eta$. For the pure model, $D_{R}=0$, we
used $\eta=0.03$ that ensures the fastest relaxation. In the presence
of RA overrelaxation leads to the energy decrease since the effective
field depends on the spin orientation. One can show that flipping
of a spin $i$ changes the energy by
\begin{equation}
\Delta E_{i}=-\frac{D_{R}}{2}\left(\Delta\mathbf{s}_{i}\cdot\mathbf{n}_{i}\right)^{2},\label{Delta_E}
\end{equation}
where $\Delta\mathbf{s}_{i}$ is the change of the spin vector $\mathbf{s}_{i}$
as the result of the transformation above. Thus for $D_{R}>0$ one
can use $\eta=0$ that leads to a relaxation much faster and deeper
in the energy than the pure spin alignment, $\eta=1$. Still, we used
$\eta=0.03$ in all cases. 

In the RA system, there are many local energy minima, and the result
depends on the initial state. Starting with the collinear state, one
arrives at more ordered states with a higher magnetization $m\equiv\left|\mathbf{m}\right|$,
where
\begin{equation}
\mathbf{m}=\frac{1}{N}\sum_{i}\mathbf{s}_{i}\label{magnetization_Def}
\end{equation}
is the average spin and $N$ is the number of spins in the system.
Using the Imry-Ma argument, one can estimate that spins are strongly
correlated within large regions of the characteristic size $R_{f}$
(ferromagnetic correlation radius) defined by 
\begin{equation}
\frac{R_{f}}{a}\sim\left(\frac{J}{D_{R}}\right)^{2/(4-d)},\label{Rf_IM}
\end{equation}
where $a$ is the lattice spacing and $d$ is the dimensionality of
the space, here $d=2$. In the case $L\gg R_{f}$, where $L$ is the
linear system's size, starting from RIC one arrives at an almost fully
disordered state, $m\ll1$. 

Starting from a random initial condition, one obtains spin states
with a very small magnetization that decreases with the system size.
However, starting from CIC one obtains a state with a significant
residual magnetization $m$ in spite of disordering due to RA. As
$D_{R}$ increases from zero, the system quickly disorders down to
$m\approx0.7$ and remains approximately constant up to $D_{R}/J\approx5$.
Then with a further increase of $D_{R}$ it gradually decreases down
to $m=0.5$ that corresponds to all spins directed along their local
anisotropy axes within a hemisphere. We denote the resulting spin
state as $\mathbf{s}_{i}^{(0)}$ and will use it to study the dynamics
of the deviations from this state in the localized modes.

The second step is performing the Metropolis Monte Carlo process at
some temperature $T$ to obtain a thermalized state. We are especially
interested in very low temperatures, such as $T/J=0.01$, for which
the thermalized state is close to the local-energy-minimum state found
in the first step. It is well known that combining Monte Carlo with
overrelaxation is very efficient and leads to much faster thermalization
than pure Monte Carlo. However, in the presence of RA overrelaxation
does not work as it does not conserve the energy, see Eq. (\ref{Delta_E}).
Here, overrelaxation can be replaced by the thermalized overrelaxation
\citep{garchu-JPhys2022} including spin flips over the part of the
effective field that does not include single-site interactions such
as RA and accepting or rejecting them with the use of the Metropolis
criterion such as in the Monte Carlo routine, whereas the energy change
arizes because of the single-site interactions. The thermodynamic
consistency of this method has been checked by computing the dynamical
spin temperature $T_{S}$ given by Eq. (9) of Ref. \citep{gar21pre}.
The values of $T_{S}$ are in a good accordance with the set temperature
$T$.

The third and most demanding step is to run the dynamical evolution
of the system by solving the dissipationless LL equation of motion,
Eq. (\ref{Larmor}) with $\alpha=0$, in a long time interval. We
are using the Butcher's fifth-order Runge-Kutta method, RK5 (see the
code, e.g., in the Appendix of Ref. \citep{gar17pre}).

\begin{figure}
\begin{centering}
\includegraphics[width=9cm]{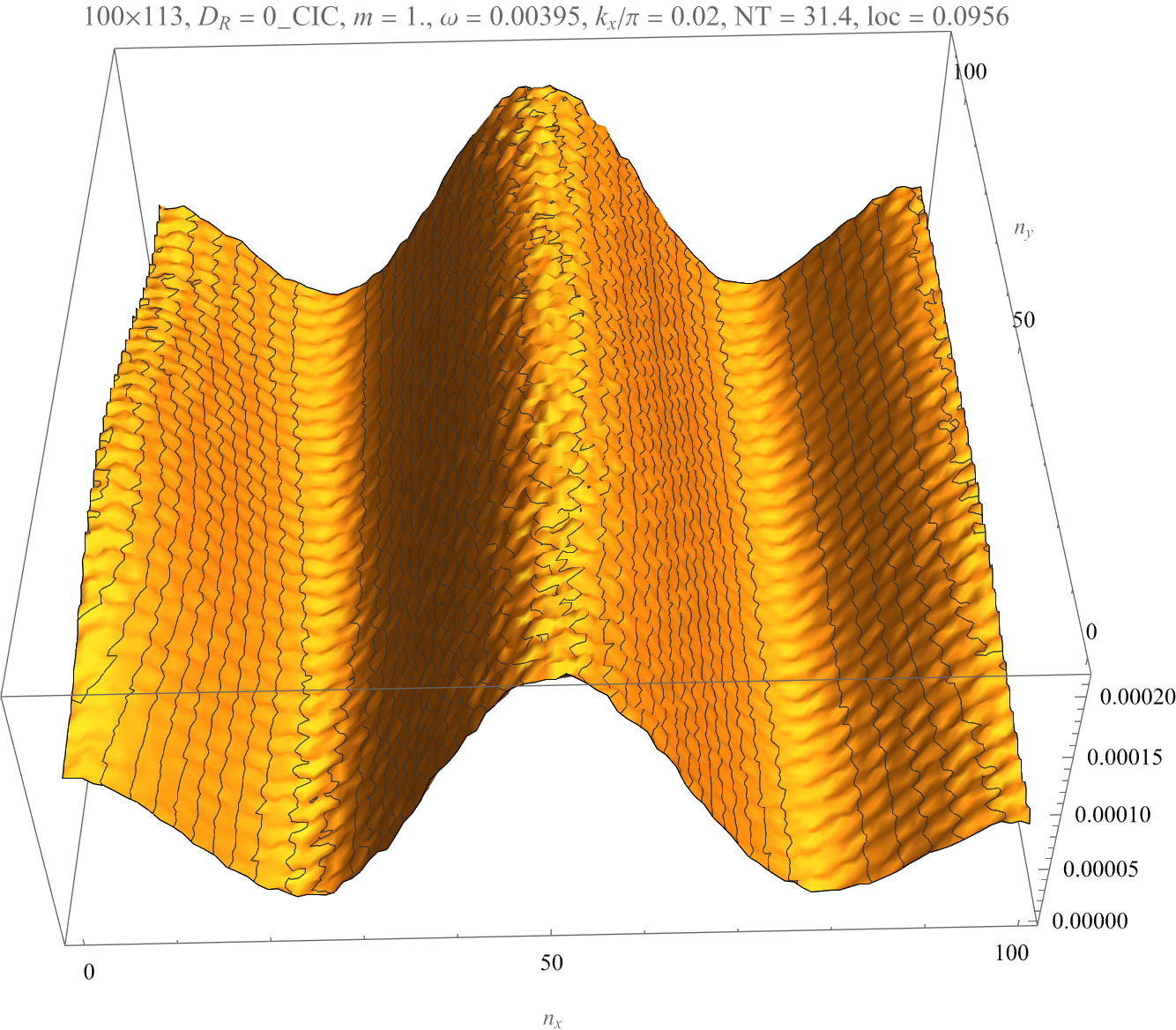}
\par\end{centering}
\begin{centering}
\includegraphics[width=9cm]{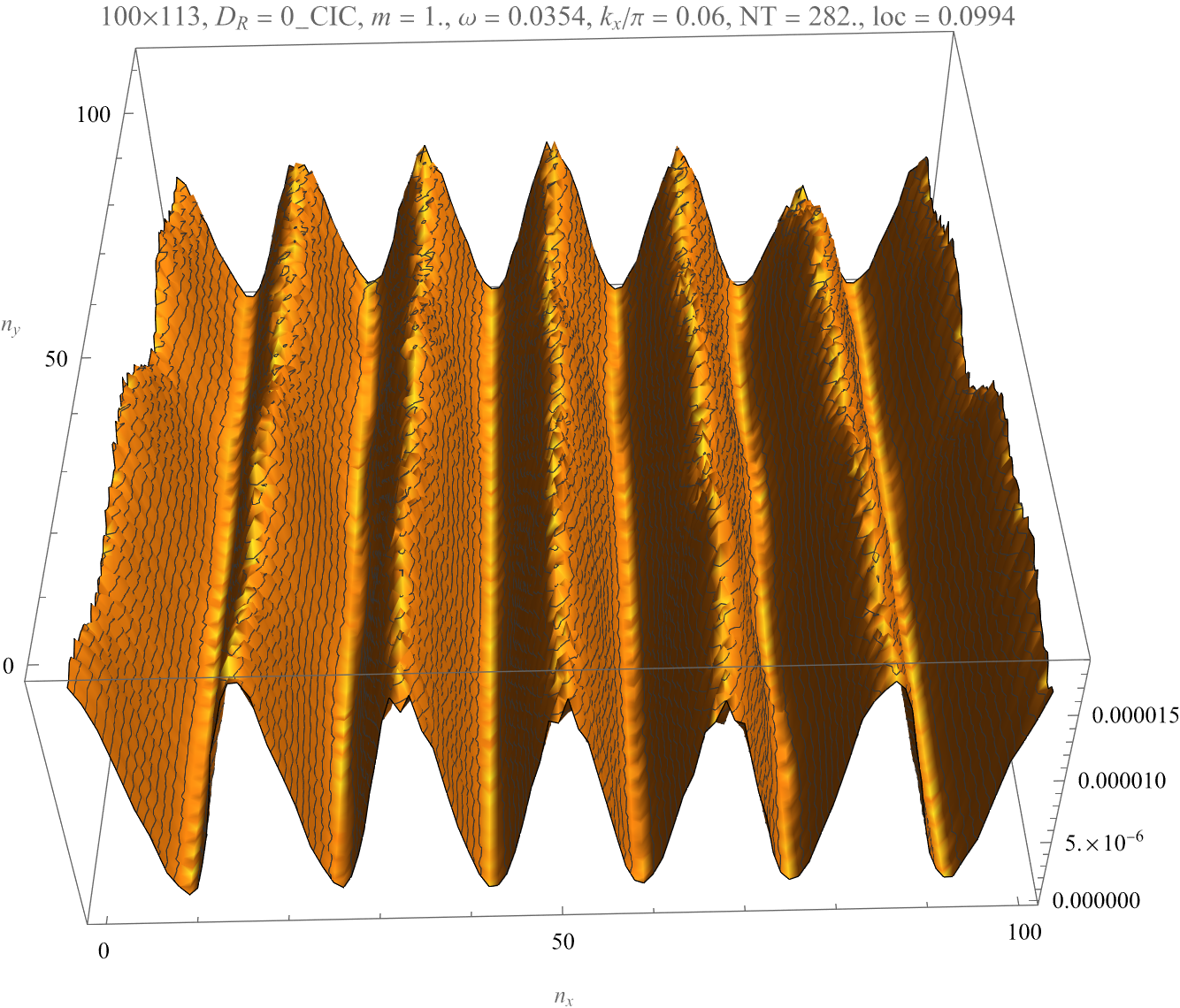}
\par\end{centering}
\caption{Spatial profiles of spin fluctuations $P_{i}(\omega)$, Eq. (\ref{Pi_Def}),
in the pure $100\times113$ system at $H=0$ and $T/J=0.01$ at different
frequencies matching those of spin-wave modes, Eq. (\ref{omega_k_Def}).
Upper panel: $k_{x}/\pi=0.02$. Lower panel: $k_{x}/\pi=0.06$.}

\label{Fig-Modes_pure}
\end{figure}

We consider the system with free boundary conditions for which the
SW eigenstates have the factorized form \citep{kacgar01pasurf}
\begin{equation}
F_{\mathbf{k}i}=f_{n_{x}k_{x}}\times f_{n_{y}k_{y}},\label{F_eigen_Def}
\end{equation}
where the site index $i$ is represented as $i=(n_{x},n_{y})$ with
$n_{x}=1,2,\ldots,N_{x}$, $n_{y}=1,2,\ldots,N_{y}$ and $N_{x}N_{y}=N$
for the rectangular-shape system. The $f$-functions have the form
of standing waves
\begin{equation}
f_{n_{\alpha}k_{\alpha}}=\sqrt{\frac{2}{1+\delta_{k_{\alpha}0}}}\cos\left[\left(n_{\alpha}-\frac{1}{2}\right)ak_{\alpha}\right],\label{f_eigen_Def}
\end{equation}
where $\alpha=x,y$ and $ak_{\alpha}=\pi n_{k_{\alpha}}/N_{\alpha}$
with $n_{k_{\alpha}}=0,1,\ldots,N_{\alpha}-1$. For a comparizon,
in a system with periodic boundary conditions (pbc) the eigenstates
are $e^{i\mathbf{k}\cdot\mathbf{r}}$ and the wave vectors are quantized
as $ak_{\alpha}=2\pi n_{k_{\alpha}}/N_{\alpha}$ with $n_{k_{\alpha}}=0,1,\ldots,N_{\alpha}-1$.
This difference is, in fact, arizing because of the degeneracy of
the wave vectors with respect to their directions in the pbc model
that makes the Brillouin zone larger and the values of $\mathbf{k}$
rarified in comparison the the fbc model. For large $N$ the fbc model
is closer to the continuous limit than the pbc model. The amplitudes
of standing waves $\mathbf{A}_{\mathbf{k}}$ can be computed as the
projection of the spin state $\mathbf{s}_{n_{x}n_{y}}$ on the eigenfunctions
set as
\begin{equation}
\mathbf{A}_{\mathbf{k}}(t)=\frac{1}{N}\sum_{n_{x}n_{y}}\mathbf{s}_{n_{x}n_{y}}(t)f_{n_{x}k_{x}}f_{n_{y}k_{y}}.\label{A_kx_Def}
\end{equation}

For $D_{R}=0$ and very low temperatures, transverse correlation functions
$G_{tr,k}(t)$ are pure sinusoidals with the frequencies
\begin{equation}
\hbar\omega_{\mathbf{k}}=4J\left\{ 1-\frac{1}{2}\left[\cos\left(ak_{y}\right)+\cos\left(ak_{y}\right)\right]\right\} .\label{omega_k_Def}
\end{equation}
 However, at nonzero temperatures they become damped because of spin-wave
processes and at elevated temperatures they become overdamped. Random
anisotropy causes damping of spin waves in the whole temperature range.
If RA is strong enough, SW become overdamped and the description of
the system's excitation modes in terms of the sinusoidal waves with
a particular wave vectors becomes inadequate. In this case, excitation
modes become localized.

\begin{figure}
\begin{centering}
\includegraphics[width=9cm]{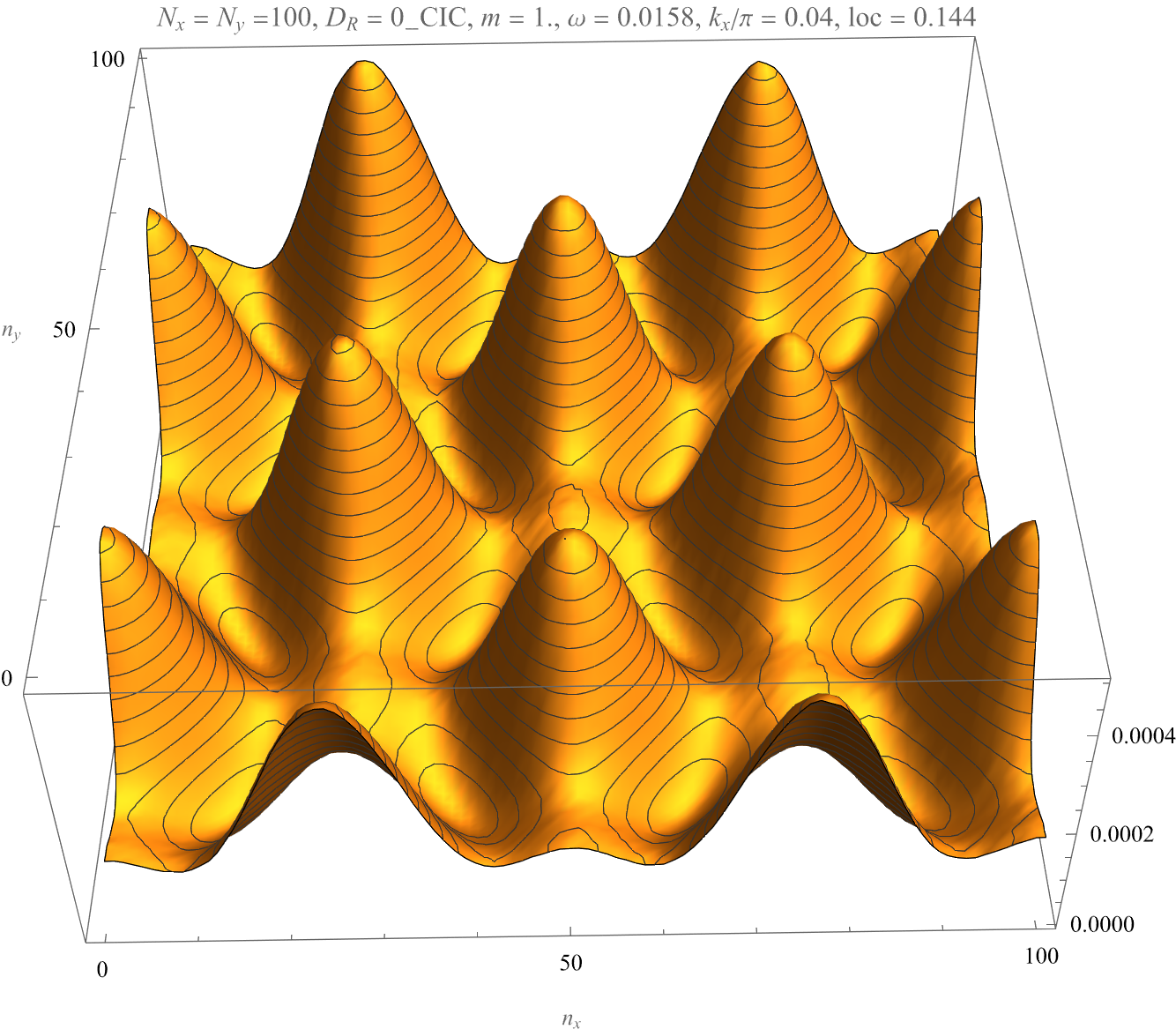}
\par\end{centering}
\begin{centering}
\includegraphics[width=9cm]{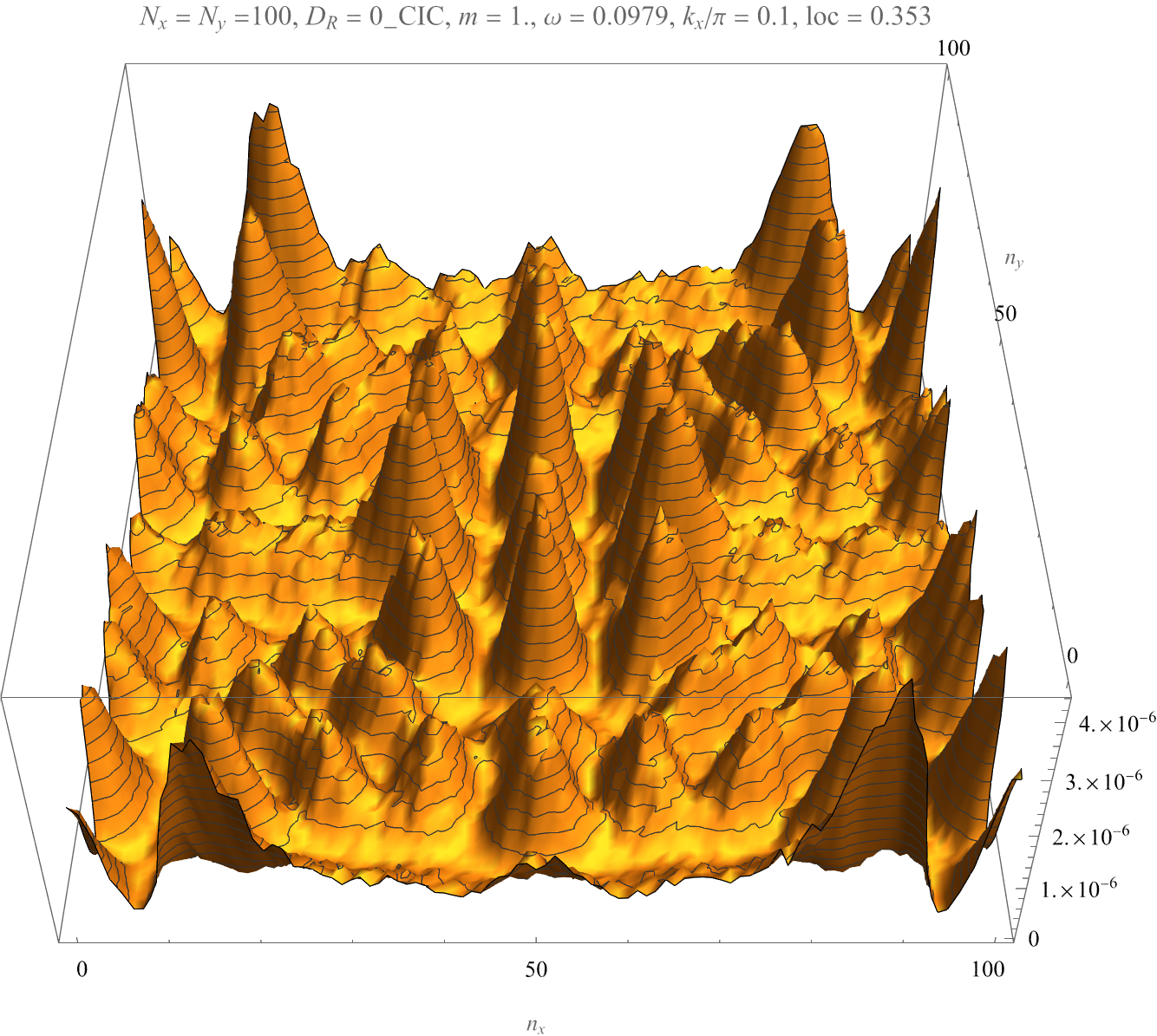}
\par\end{centering}
\caption{Spatial profiles of spin fluctuations $P_{i}(\omega)$, Eq. (\ref{Pi_Def}),
in the pure system at $H=0$ and $T/J=0.01$ at different frequencies
matching those of spin-wave modes. Upper panel: $k/\pi=0.04$. Lower
panel: $k/\pi=0.1$.}

\label{Fig-Modes_pure_superposition}
\end{figure}

To study extended and localized modes, we use the following way of
processing the dynamical evolution result $\mathbf{s}_{i}(t)$. For
a very low temperature, such as $T/J=0.01$, we perform a Fourier
transformation of the deviation of the spins from the local energy-minimum
state, $\delta\mathbf{s}_{i}(t)\equiv\mathbf{s}_{i}(t)-\mathbf{s}_{i}^{(0)}$
with a set of frequencies, say, the frequencies given by Eq. (\ref{omega_k_Def})
for the chosen set of $\mathbf{k}$:
\begin{equation}
\delta\mathbf{\tilde{s}}_{i}(\omega)=\frac{1}{\sqrt{t}}\intop_{0}^{t}dt'e^{i\omega t'}\delta\mathbf{s}_{i}(t').\label{s_Fourier}
\end{equation}
The spin deviations $\delta\mathbf{s}_{i}$ are expected to precess
around $\mathbf{s}_{i}^{(0)}$. Thus it is convenient to introduce
local axis $\mathbf{e}_{iz'}=\mathbf{s}_{i}^{(0)}$ and the two perpendicular
axes $\mathbf{e}_{ix'}$ and $\mathbf{e}_{iy'}$. Fixing $\mathbf{e}_{iy'}$
in the $xy$ plane and choosing $\mathbf{e}_{ix'}$ perpendicular
to both $\mathbf{e}_{iz'}$ and $\mathbf{e}_{iy'}$, one obtains
\begin{equation}
\mathbf{e}_{iy'}=\frac{\mathbf{s}_{i}^{(0)}\times\mathbf{e}_{z}}{\left|\mathbf{s}_{i}^{(0)}\times\mathbf{e}_{z}\right|},\qquad\mathbf{e}_{ix'}=\mathbf{e}_{iy'}\times\mathbf{s}_{i}^{(0)}.\label{expr_eypr_def}
\end{equation}
In the case $\mathbf{s}_{i}^{(0)}\times\mathbf{e}_{z}=0$ we just
choose $\mathbf{e}_{ix'}=\mathbf{e}_{x}$ and $\mathbf{e}_{iy'}=\mathbf{e}_{y}$. 

Then we introduce the complex quantity 
\begin{equation}
p_{i}(\omega)\equiv\delta\mathbf{\tilde{s}}_{i}(\omega)\cdot\left(\mathbf{e}_{ix'}+i\mathbf{e}_{iy'}\right).\label{pi_Def}
\end{equation}
 Further, we introduce the real quantity
\begin{equation}
P_{i}(\omega)\equiv2\delta\mathbf{\tilde{s}}_{i}(\omega)\cdot\delta\mathbf{\tilde{s}}_{i}^{*}(\omega),\label{Pi_Def}
\end{equation}
that characterizes the spatial profile of spin fluctuations at the
frequency $\omega$. The function $P_{i}(\omega)$ has an apreciable
value if the frequency $\omega$ coincides with one of the system's
excitation frequencies $\omega_{\mu}$. If the modes are localized,
$P_{i}(\omega)$ consists of peaks corresponding to these modes.

\begin{figure}
\begin{centering}
\includegraphics[width=9cm]{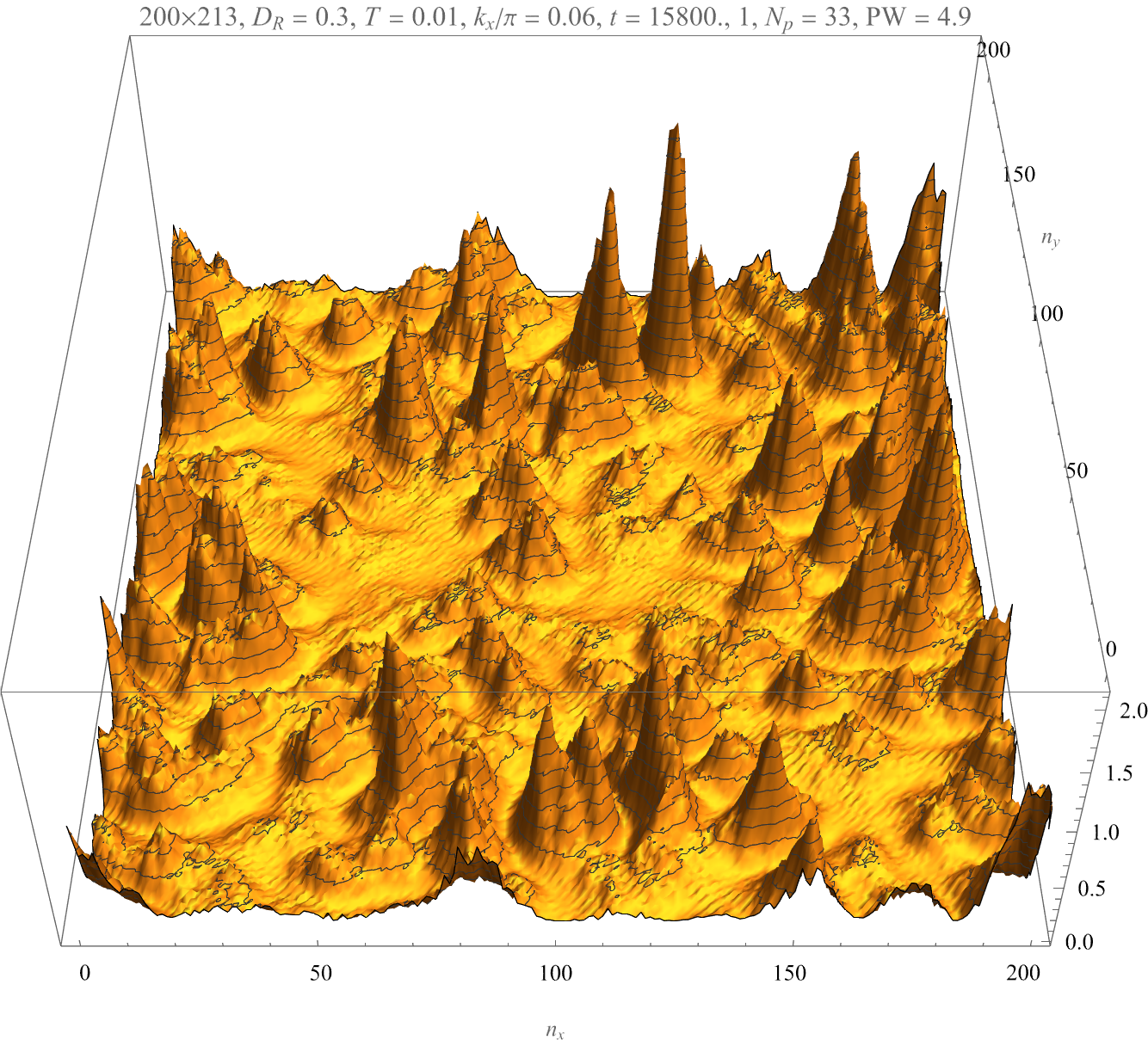}
\par\end{centering}
\begin{centering}
\includegraphics[width=9cm]{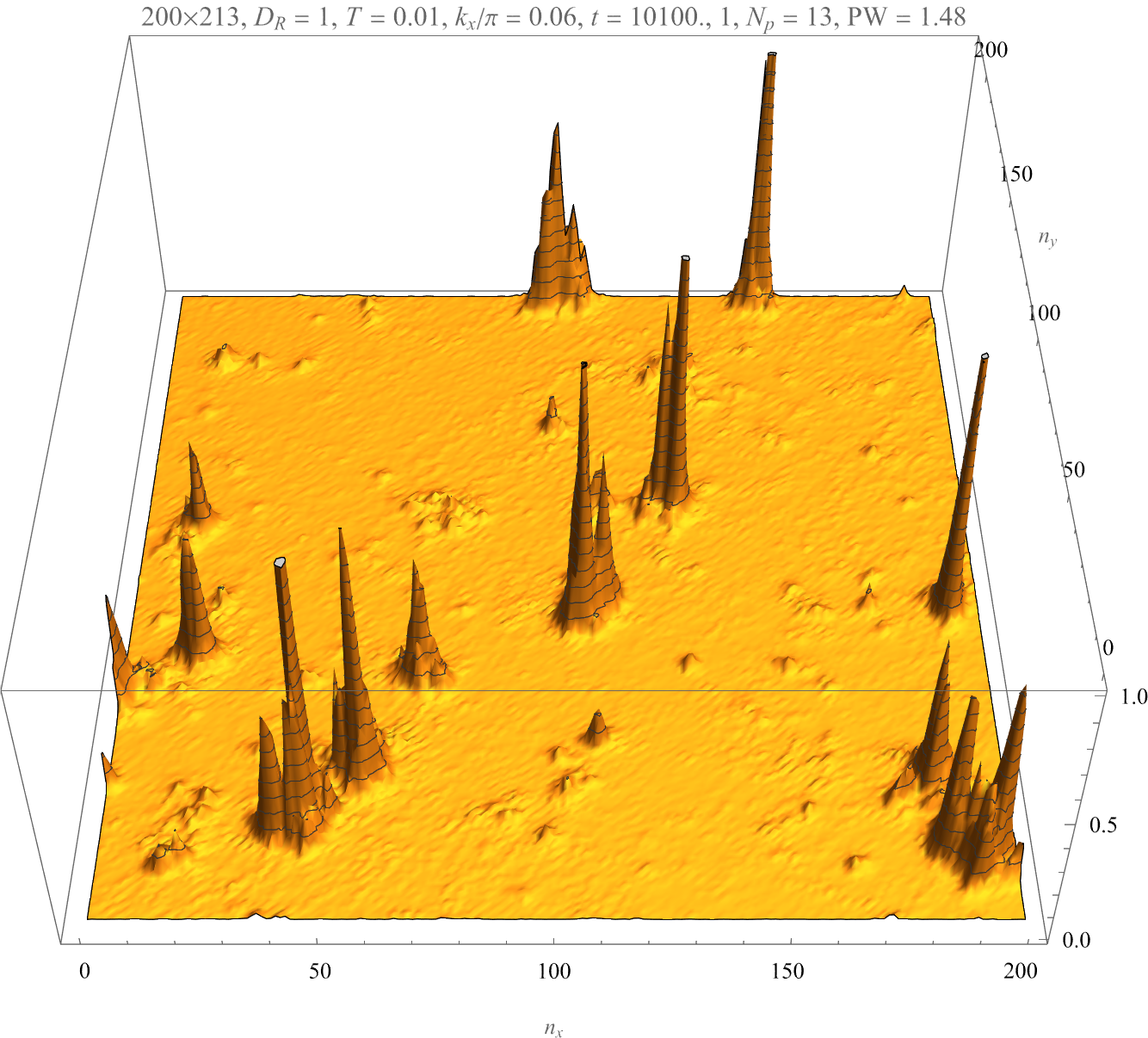}
\par\end{centering}
\caption{Normalized spatial profiles of spin fluctuations $P_{i}(\omega)$,
Eq. (\ref{Pi_Def}), in the $200\times2130$ RA system with $D_{R}/J=0.3$
(upper panel) and $D_{R}/J=1$ (lower panel) at $H=0$ and $T/J=0.01$
at one of the frequencies matching those of spin-wave modes, Eq. (\ref{omega_k_Def}):
$\hbar\omega/J=0.0354$ corresponding to $k/\pi=0.06$.}

\label{Fig-Modes_100x100_DR=00003D1}
\end{figure}

Another useful quantity is the projection of the excitation mode with
the frequency $\omega$ on the eigenmode $\mathbf{k}$ 
\begin{equation}
Q_{\mathbf{k}}(\omega)\equiv\left(\frac{1}{N}\sum_{i}\delta\mathbf{\tilde{s}}_{i}(\omega)F_{\mathbf{k}i}\right)\cdot\left(\frac{1}{N}\sum_{i}\delta\mathbf{\tilde{s}}_{i}^{*}(\omega)F_{\mathbf{k}i}\right).\label{Q_omega_Def}
\end{equation}
Especially interesting is $Q_{0}(\omega)$, the coupling of the mode
$\omega$ to the uniform time-dependent magnetic field of such frequency.
The latter is related to the absorption of microwave energy by this
magnetic system. In this case one has
\begin{equation}
Q_{0}(\omega)\equiv\frac{1}{N^{2}}\sum_{ii'}\delta\mathbf{\tilde{s}}_{i}(\omega)\cdot\delta\mathbf{\tilde{s}}_{i'}^{*}(\omega)=\delta\tilde{\mathbf{m}}(\omega)\cdot\delta\tilde{\mathbf{m}}^{*}(\omega),\label{Q0_omega_def}
\end{equation}
where $\mathbf{m}$ is the average spin (magnetization), see Eq. (\ref{magnetization_Def}).
Further,
\begin{equation}
\delta\tilde{\mathbf{m}}(\omega)\cdot\delta\tilde{\mathbf{m}}^{*}(\omega)=\frac{1}{t}\intop_{0}^{t}dt'dt''e^{i\omega\left(t'-t''\right)}\delta\mathbf{m}(t')\cdot\delta\mathbf{m}(t'').
\end{equation}
Expressing the product $\delta\mathbf{m}(t')\cdot\delta\mathbf{m}(t'')$
through the correlation function $G(\tau)$ depending on the time
difference $\tau\equiv t'-t''$, one obtains
\begin{equation}
Q_{0}(\omega)=\int_{-\infty}^{\infty}d\tau e^{i\omega\tau}G(\tau)=\tilde{G}(\omega).\label{Q0_via_G}
\end{equation}
The absorbed power $P_{\mathrm{abs}}(\omega)$ of the microwave radiation
per spin can be expressed with the help of the FDT as \citep{garchu-JPhys2022}
\begin{equation}
\frac{P_{\mathrm{abs}}(\omega)}{h_{0}^{2}}=\frac{\omega^{2}N}{12k_{B}T}\tilde{G}(\omega)\label{Pabs}
\end{equation}

\section{Expansion over excitation modes}

\label{sec:Modeling-of-excitation}

Expanding $\delta\mathbf{s}_{i}(t)$ over the complete set of excitation
modes, one can write
\begin{equation}
\delta\mathbf{s}_{i}(t)=\sum_{\mu}A_{\mu i}\left[\mathbf{e}_{ix'}\cos\left(\omega_{\mu}t+\phi_{\mu}\right)-\mathbf{e}_{iy'}\sin\left(\omega_{\mu}t+\phi_{\mu}\right)\right],\label{delta_s_via_local_modes}
\end{equation}
where $A_{\mu i}$ is localized in space with the magnitude depending
on the degree of excitation of the system. The spins are precessing
clockwise in the Landau-Lifshitz equation, hence the minus sign in
the second term. In RA systems, the magnetization is changing at the
distances $R_{f}$, and if the modes are localized at smaller distances,
the representation above should be valid. 

Keeping only the slowly-decaying terms with the frequency difference
$\omega-\omega_{\mu}$ and discarding the terms with $\omega+\omega_{\mu}$
in Eq. (\ref{s_Fourier}), one obtains
\begin{eqnarray}
\delta\mathbf{\tilde{s}}_{i}(\omega) & = & \sum_{\mu}A_{\mu i}\left(\mathbf{e}_{ix'}-i\mathbf{e}_{iy'}\right)\nonumber \\
 & \times & e^{i\left(\omega-\omega_{\mu}\right)t/2-i\phi_{\mu}}\frac{\sin\left[\left(\omega-\omega_{\mu}\right)t/2\right]}{\left(\omega-\omega_{\mu}\right)\sqrt{t}}.\label{s_Fourier_result}
\end{eqnarray}
At large times only the terms with $t\left(\omega-\omega_{\mu}\right)\lesssim1$
survive in this formula, so that the complex exponential is approximately
a constant phase factor $e^{-i\phi_{\mu}}$. The latter depend on
how the mode $\mu$ was excited and are unique for each excitation
mode. One can express the function $p_{i}(\omega)$ in Eq. (\ref{pi_Def})
in terms of the local modes as follows:
\begin{equation}
p_{i}(\omega)=\sum_{\mu}A_{\mu i}e^{i\left(\omega-\omega_{\mu}\right)t/2-i\phi_{\mu}}\frac{2\sin\left[\left(\omega-\omega_{\mu}\right)t/2\right]}{\left(\omega-\omega_{\mu}\right)\sqrt{t}}.\label{pi_result}
\end{equation}
 As mention above, the quantity $p_{i}(\omega)$ is similar to the
quantum wave function $\Psi$ of the system that is a superposition
of all eigenfunctions $\mu$ of a given energy $E_{\mu}=\hbar\omega_{\mu}$.
One can plot $\left|\varPsi\right|^{2}$ on the latttice that results
in a system of standing waves for the pure system and peaks correcponding
to local modes for the RA system. The phase of the $p_{i}(\omega)$
as the function of the position on the lattice $i$ should be nearly
constant in the region of a particular localized mode. For instance,
looking at the correlation of the phases of two peaks close to each
other, one can find out whether these peaks belong to the same or
to different local modes. 

Let us now calculate $P_{i}(\omega)$ defined by Eq. (\ref{Pi_Def}).
Using Eq. (\ref{delta_s_via_local_modes}) and neglecting cross-terms
with different values of $\mu$ and different phases, as well as fast-decaying
terms with $\omega+\omega_{\mu}$, one obtains
\begin{equation}
P_{i}(\omega)=\sum_{\mu}A_{\mu i}^{2}\frac{4\sin^{2}\left[\left(\omega-\omega_{\mu}\right)t/2\right]}{\left(\omega-\omega_{\mu}\right)^{2}t}\label{Pi_omega_analytical}
\end{equation}
that is obviously related to $p_{i}(\omega)$ above. The terms of
this expression decrease slowly in time if $\omega$ is close to $\omega_{\mu}$
and are equal to $1/2$ for $\omega=\omega_{\mu}$. The quantity $P_{i}(\omega)$
selects all modes with frequency $\omega_{\mu}=\omega$. The longer
is the integration time $t$, the sharper is the selection. In the
limit $t\rightarrow\infty$ the terms Eq. (\ref{Pi_omega_analytical})
this becomes
\begin{equation}
P_{i}(\omega)=2\pi\sum_{\mu}A_{\mu i}^{2}\delta\left(\omega-\omega_{\mu}\right).\label{P_omega_asymp_t}
\end{equation}
 In the pure system, excitation modes' profile $P_{i}(\omega)$ is
a superposition of standing waves. In the presence of RA, $P_{i}(\omega)$
consists of peaks corresponding to different localized modes. Numerical
calculations indeed show the behavior $P_{i}(\omega)\propto1/t$.

For $Q_{0}(\omega)$ defined by Eq. (\ref{Q0_omega_def}), using the
representation of spin deviations in terms of local modes, Eq. (\ref{delta_s_via_local_modes}),
one can use
\begin{equation}
\delta\mathbf{m}(t)=\frac{1}{N}\sum_{\mu i}A_{\mu i}\left[\mathbf{e}_{ix'}\cos\left(\omega_{\mu}t+\phi_{\mu}\right)-\mathbf{e}_{iy'}\sin\left(\omega_{\mu}t+\phi_{\mu}\right)\right].
\end{equation}
Discarding cross-terms with $\mu\neq\mu'$ and assuming $\mathbf{e}_{i'x'}\cdot\mathbf{e}_{i''x'}\cong\mathbf{e}_{i'y'}\cdot\mathbf{e}_{i''y'}\cong1$
and $\mathbf{e}_{i'x'}\cdot\mathbf{e}_{i''y'}\cong\mathbf{e}_{i'y'}\cdot\mathbf{e}_{i''x'}\cong0$,
one obtains
\begin{equation}
Q_{0}(\omega)=\frac{1}{t}\intop_{0}^{t}dt'dt''e^{i\omega\left(t'-t''\right)}\frac{1}{N^{2}}\sum_{\mu}\bar{A}_{\mu}^{2}\cos\left[\omega_{\mu}\left(t'-t''\right)\right],\label{Q0_via_tt}
\end{equation}
where 
\begin{equation}
\bar{A}_{\mu}\equiv\sum_{i}A_{\mu i}\label{A_mu_avr}
\end{equation}
is the spatial average of the excitation-mode amplitude. One can see
that only modes with a nonzero spatial average contribute to the power
absorption as they couple to the uniform time-dependent external field.
Integrating over the difference time in Eq. (\ref{Q0_via_tt}) as
above, one obtains
\begin{equation}
Q_{0}(\omega)=\frac{1}{N^{2}}\sum_{\mu}\bar{A}_{\mu}^{2}\intop_{-t}^{t}d\tau e^{i\omega\tau}\cos\left(\omega_{\mu}\tau\right).
\end{equation}
At large times this becomes
\begin{equation}
Q_{0}(\omega)=\frac{\pi}{2}\frac{1}{N^{2}}\sum_{\mu}\bar{A}_{\mu}^{2}\delta\left(\omega-\omega_{\mu}\right).\label{Q0_asymp_t}
\end{equation}
Comparing this with Eq. (\ref{Q0_via_G}), one obtains the Fourier
transform of the time correlation function of the magnetization as
\begin{equation}
\tilde{G}(\omega)=Q_{0}(\omega)=\frac{\pi}{2}\frac{1}{N^{2}}\sum_{\mu}\bar{A}_{\mu}^{2}\delta\left(\omega-\omega_{\mu}\right).\label{G_omega_via_A_mu}
\end{equation}
This quantity defines the absorbed power given by Eq. (\ref{Pabs}). 

\subsection{Characterization of the localized modes}

\label{subsec:Characterization-of-the}One of the quantities characterizing
excitation modes is the average peak width $\delta$ of the lanscape
of $P_{i}\equiv P_{n_{x}n_{y}}$. To estimate it, one can use two
different methods. The first method is more ``pedestrian'' and directly
computes the averaged half-width of the peaks at their average half-height.
To find the average height of the peaks, we partitioned the system
into $10\times10$ blocks, found the maximal values of $P_{n_{x}n_{y}}$
in each block and then averaged these values. Then the total number
of sites having the height exceeding the half of the maximal height
was computed. The result, in lattice units, is the area of all peaks
$A_{p}=N_{p}\pi\delta^{2}$, where $N_{p}$ is the number of peaks
and $\delta$ is the peak width at the half-height. Computing the
number of peaks $N_{p}$ is a separate problem. One cannot just find
all maxima of the function $P_{n_{x}n_{y}}$ because in a random system
there are lots of minor peaks on the slope of a major peak. Taking
all them into account leads to a strong underestimation of the average
peak width that should be defined by the major peaks only. The solution
is to consider the ``islands'' of $P_{n_{x}n_{y}}$that are above
the half of the average peak height $B$ as major peaks. Counting
the number of these islands is a standard problem of \textit{connected-component
labeling} that arizes, in particular, in image recognition. So, we
used one of the available algorithms to compute $N_{p}$, after which
we found $\delta=\sqrt{A_{p}/(\pi N_{p})}$.

Another method is more sophisticated and uses summation of different
powers $m$ and $n$ of $P_{n_{x}n_{y}}$ to extract the peaks' height
and the area of all peaks. It is assumed that $P_{n_{x}n_{y}}$ is
a collection of Gaussian peaks $f(r)=B\exp\left(-r^{2}/\delta^{2}\right)$
, Lorentzian peaks $f(r)=B/\left(1+r^{2}/\delta^{2}\right)$, or any
bell-like functions. Assuming that there are $N_{p}$ such non-overlapping
peaks, one can define 
\begin{equation}
W_{m}\equiv\intop_{0}^{\infty}d^{2}rf^{m}(r)dr=N_{p}B^{m}\delta^{2}q_{m},\label{Wm_Def}
\end{equation}
where $q_{m}$ is a number: $q_{m}=\pi/m$ for the Gaussian function
and $q_{m}=\pi/(m-1)$ for the Lorentzian function. One can see that
$W_{m}$ depends on the peak height $B$ and the area of all peaks
$A_{p}=N_{p}\pi\delta^{2}$. For the Lorentzian function, $\delta$
is exactly the half width at the half-height of the peak, thus we
will use this function in interpreting the results. Having two such
integrals for different values of $m$ allows to find the peaks' height
$B$ and the area of all peaks $A_{p}$ as
\begin{equation}
B=\left(\frac{\widetilde{W}_{m}}{\widetilde{W}_{n}}\right)^{1/(m-n)},\quad A_{p}=\pi\frac{\widetilde{W}_{n}^{m/\left(m-n\right)}}{\widetilde{W}_{m}^{n/\left(m-n\right)}},\label{Peaks_height_area}
\end{equation}
where $\widetilde{W}_{m}\equiv W_{m}/q_{m}$ etc. Applying this to
the problem at hands, one computes $W_{m}=\sum_{n_{x}n_{y}}P_{n_{x}n_{y}}^{m}$
and $W_{n}=\sum_{n_{x}n_{y}}P_{n_{x}n_{y}}^{n}$ from which one finds
the average peak height and peaks' area using the formulas above.
The powers $m$ and $n$ have to be larger than one to ensure the
convergence at large distances from the peaks and to suppress the
contribution of the low-level noise background. To find the number
of peaks, one computes the number of ``islands'' in which $P_{n_{x}n_{y}}>B/2$
with the help of the same \textit{connected-component labeling} routine.
Then the peak width is defined as above, $\delta=\sqrt{A_{p}/(\pi N_{p})}$.
As the shape of the peaks is unknown, the first and simpler method
of measuring the peaks' width is preferable. The second method is
used only to check the consistency of the results.

One more parameter characterizing locallization is based on symmetry.
Increasing randomness leads to the transformation of symmetric patterns
of standing waves in $P(\omega)$, see Figs. \ref{Fig-Modes_pure}
and \ref{Fig-Modes_pure_superposition} to random patterns typical
for localized excitations, see Fig. \ref{Fig-Modes_100x100_DR=00003D1}.
Here we try to find computable quantities allowing to trace the transition
from extended to localized states. One idea is that symmetric patterns
can be reduced by antisymmetrization with respect to the center of
the system while asymmetric patterns cannot. Using the compound index
$i=(n_{x},n_{y})$ in the square lattice, for $P_{n_{x},n_{y}}(\omega)$
with $n_{x}=1,2,\ldots,N_{x}$ and $n_{y}=1,2,\ldots,N_{y}$ one can
define 
\begin{equation}
\widetilde{P}_{n_{x},n_{y}}\equiv\left|P_{n_{x},n_{y}}-P_{N_{x}-n_{x},n_{y}}\right|.
\end{equation}
For a symmetric pattern, $\widetilde{P}_{n_{x},n_{y}}$ will be zero,
while for a collection of random peaks it nearly doubles. Further,
one can define 
\begin{equation}
\widetilde{\widetilde{P}}_{n_{x},n_{y}}\equiv\frac{1}{2}\left|\widetilde{P}_{n_{x},n_{y}}-\widetilde{P}_{n_{x},N_{y}-i_{y}}\right|
\end{equation}
to further reduce the pattern and define the localization parameter
\begin{equation}
\zeta=\sum_{n_{x},n_{y}}\widetilde{\widetilde{P}}_{n_{x},n_{y}}/\sum_{n_{x},n_{y}}P_{n_{x},n_{y}}
\end{equation}
that changes between nearly zero for $D_{R}=0$ and nearly two for
$D_{R}\sim J$. In fact, it is better to use the minimal value of
the quantity defined above and that with reflection of $i_{y}$ and
then of $i_{x}$. 

\subsection{Localization and power absorption}

Localization of excitation modes in a magnetic system leads to power
absorption in a broad range of frequencies. In a pure system, there
is the uniform mode that is the only mode coupled to the microwave
field that has $\mathbf{k=0}$. All other eigenmodes of the system
are orthogonal to the uniform mode and thus they cannot absorb microwave
energy. Random anisotropy breaks translational invariance of the system,
thus plane waves, including the uniform mode, are no longer eigenmodes.
Actual eigenmodes all have a nonzero projection on the uniform mode
and thus they are coupled to the microwave field and can absorb its
energy. A sufficiently strong RA leads to localization of excitation
modes that are described by the peaks of $P_{i}(\omega)$, as shown,
e.g., in Fig. \ref{Fig-Modes_100x100_DR=00003D1}. The integrals over
the localized eigenmode amplitudes $\bar{A}_{\mu}$ in Eq. (\ref{A_mu_avr})
that enter the Fourier-transform of the time correlation function
of magnetization components, $\tilde{G}(\omega)$, are, in general,
nonzero and can be estimated via the peak parameters introduced in
Sec. \ref{subsec:Characterization-of-the}. As an estimation, one
can use
\begin{equation}
\bar{A}_{\mu}^{2}\equiv\left(\sum_{i}A_{\mu i}\right)^{2}\sim\delta^{2}\sum_{i}A_{\mu i}^{2},
\end{equation}
where $\delta$ is the average peak width in lattice units. This allows
one to relate $Q_{0}(\omega)$ in Eq. (\ref{Q0_asymp_t}) with $P_{i}(\omega)$
in Eq. (\ref{P_omega_asymp_t}) and obtain
\begin{equation}
\tilde{G}(\omega)=Q_{0}(\omega)\sim\frac{\delta^{2}}{N^{2}}\sum_{i}P_{i}(\omega)
\end{equation}
for $\tilde{G}(\omega)$ that enters the formula for the microwave
absorption, Eq. (\ref{Pabs}). The sum in this formula does not depend
on the time and scales with the system size $N$, so that $\tilde{G}(\omega)\propto1/N$.
This ensures that the absorbed power per spin in Eq. (\ref{Pabs})
does not depend on the system size. Further, one can relate the power
absorption to the average peaks height $B$ and the number of peaks
$N_{p}$ using Eqs. (\ref{Wm_Def}) and (\ref{Peaks_height_area})
with $m=1$. This yields $\sum_{i}P_{i}(\omega)\sim B\delta^{2}N_{p}$
that results in
\begin{equation}
\tilde{G}(\omega)\sim\frac{N_{p}}{N^{2}}B\delta^{4}.
\end{equation}
After that the absorbed power, Eq. (\ref{Pabs}), becomes 
\begin{equation}
\frac{P_{\mathrm{abs}}(\omega)}{h_{0}^{2}}=\frac{\omega^{2}N}{12k_{B}T}\tilde{G}(\omega)=\frac{N_{p}}{N}\frac{\omega^{2}B\delta^{4}}{12k_{B}T}.\label{Pabs-1}
\end{equation}
 One can see that power absorption depends on the peak width $\delta$,
the ratio of the peak height to the temperature $B/T$, and peaks'
concentration $N_{p}/N$, all at the frequency $\omega$. 

\begin{figure}
\begin{centering}
\includegraphics[width=9cm]{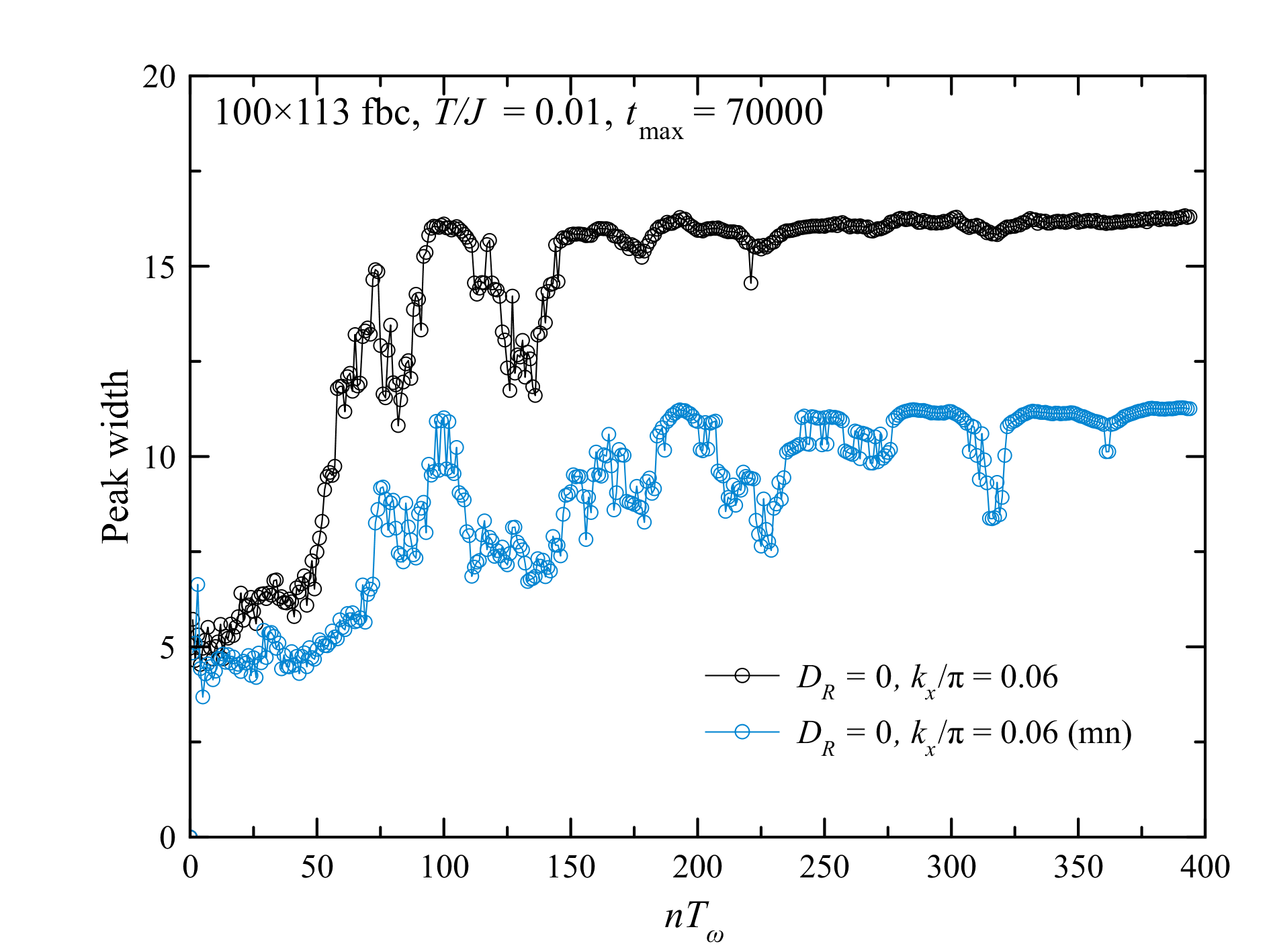}
\par\end{centering}
\begin{centering}
\includegraphics[width=9cm]{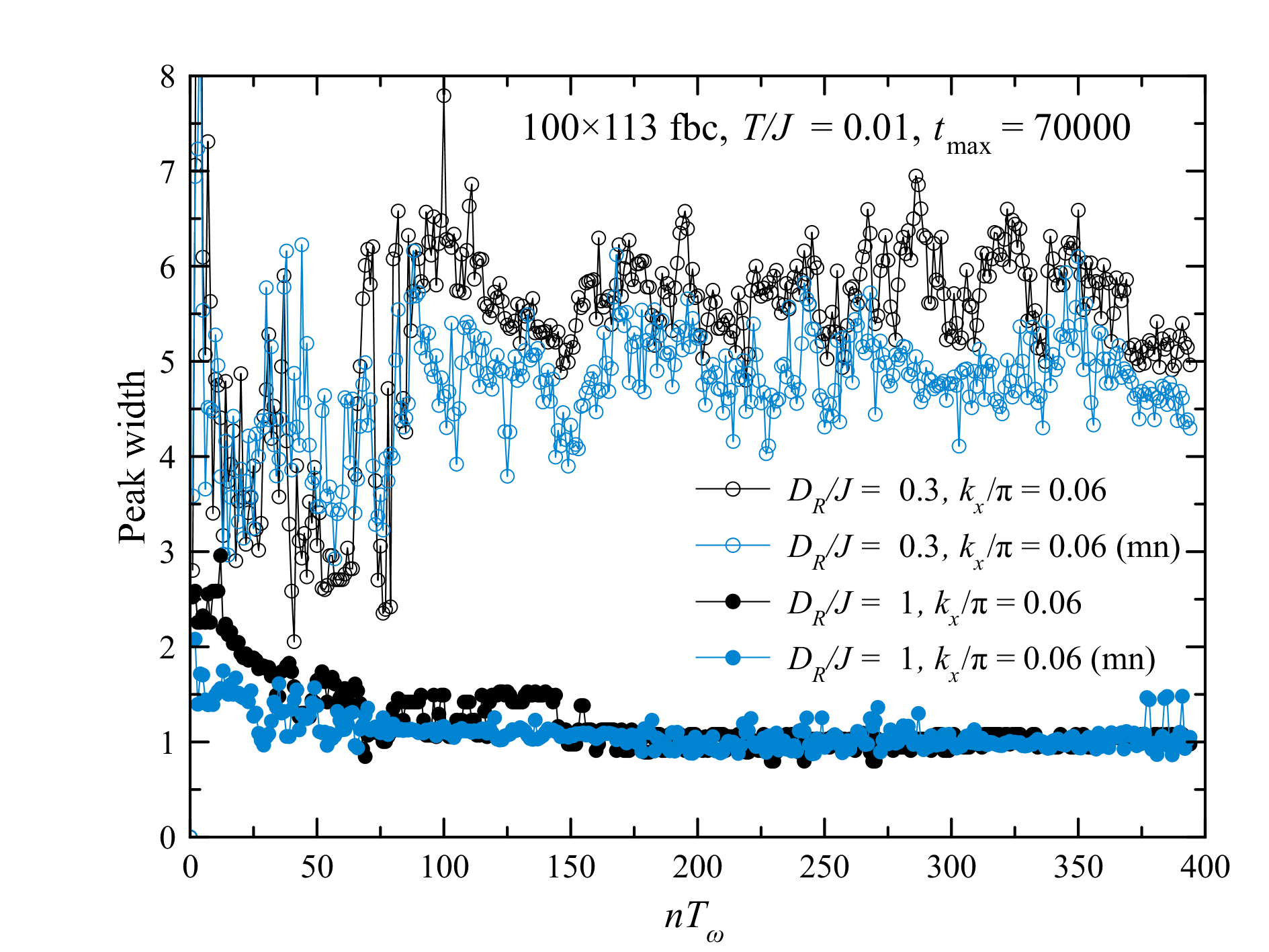}
\par\end{centering}
\caption{The average peak width vs the number of periods of the mode frequency
corresponding to the pure $k_{x}/\pi=0.06$ mode measured by two methods.
Results obtained by the second method are labeled with \textquotedblleft (mn)\textquotedblright .
Upper panel: Pure system, $D_{R}=0$. Lower panel: $D_{R}/J=0.3$
and 1.}

\label{Fig-PeakWidth_vs_nT0}
\end{figure}
\begin{figure}
\begin{centering}
\includegraphics[width=9cm]{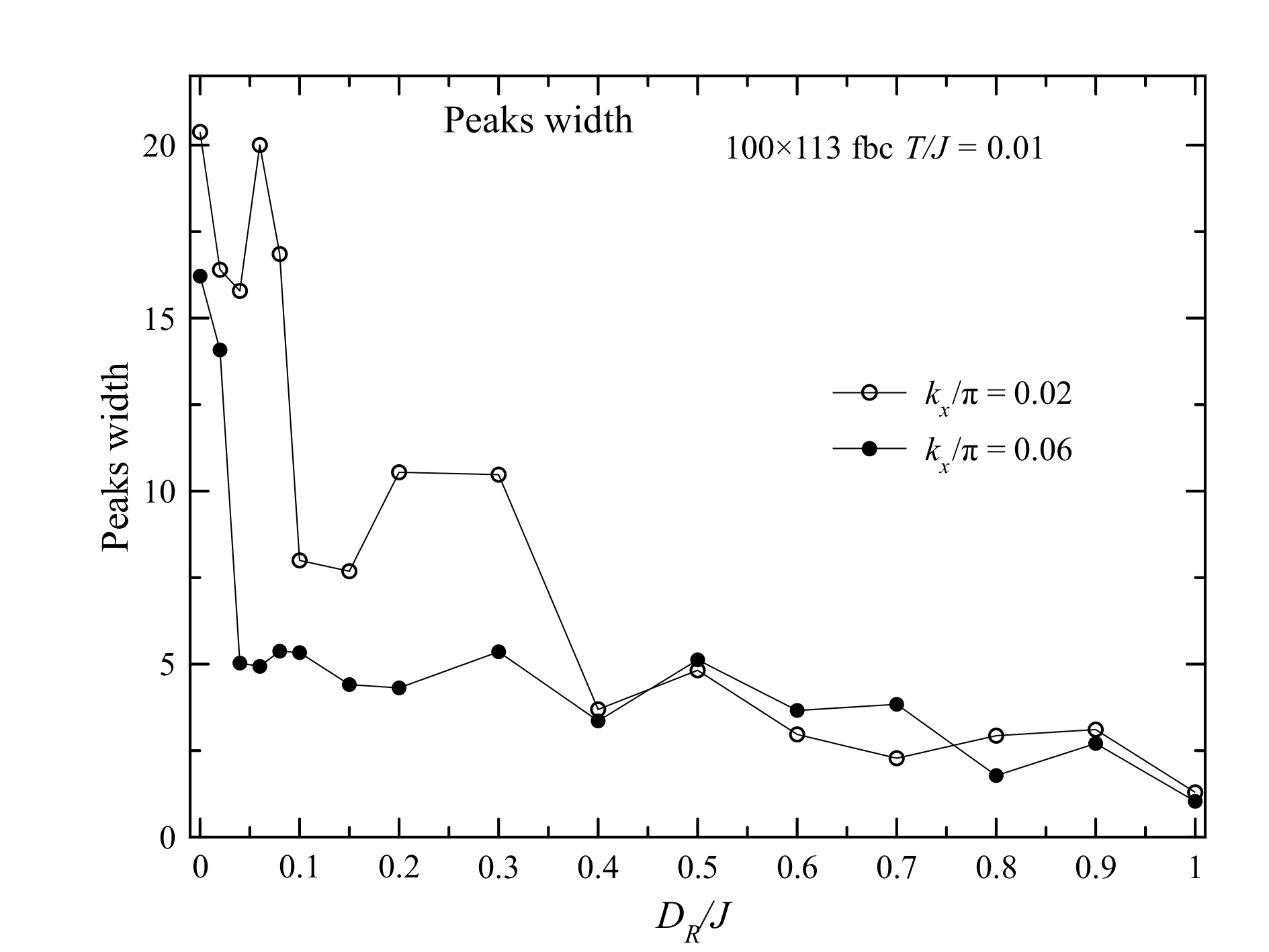}
\par\end{centering}
\caption{The average peak width $\delta$ vs $D_{R}/J$ for a system of the
size $100\times113$ and the frequencies corresponding to the pure
$k_{x}/\pi=0.02$ and 0.06 modes.}

\label{Fig-Peaks_widths_vs_DR}
\end{figure}

\begin{figure}
\begin{centering}
\includegraphics[width=9cm]{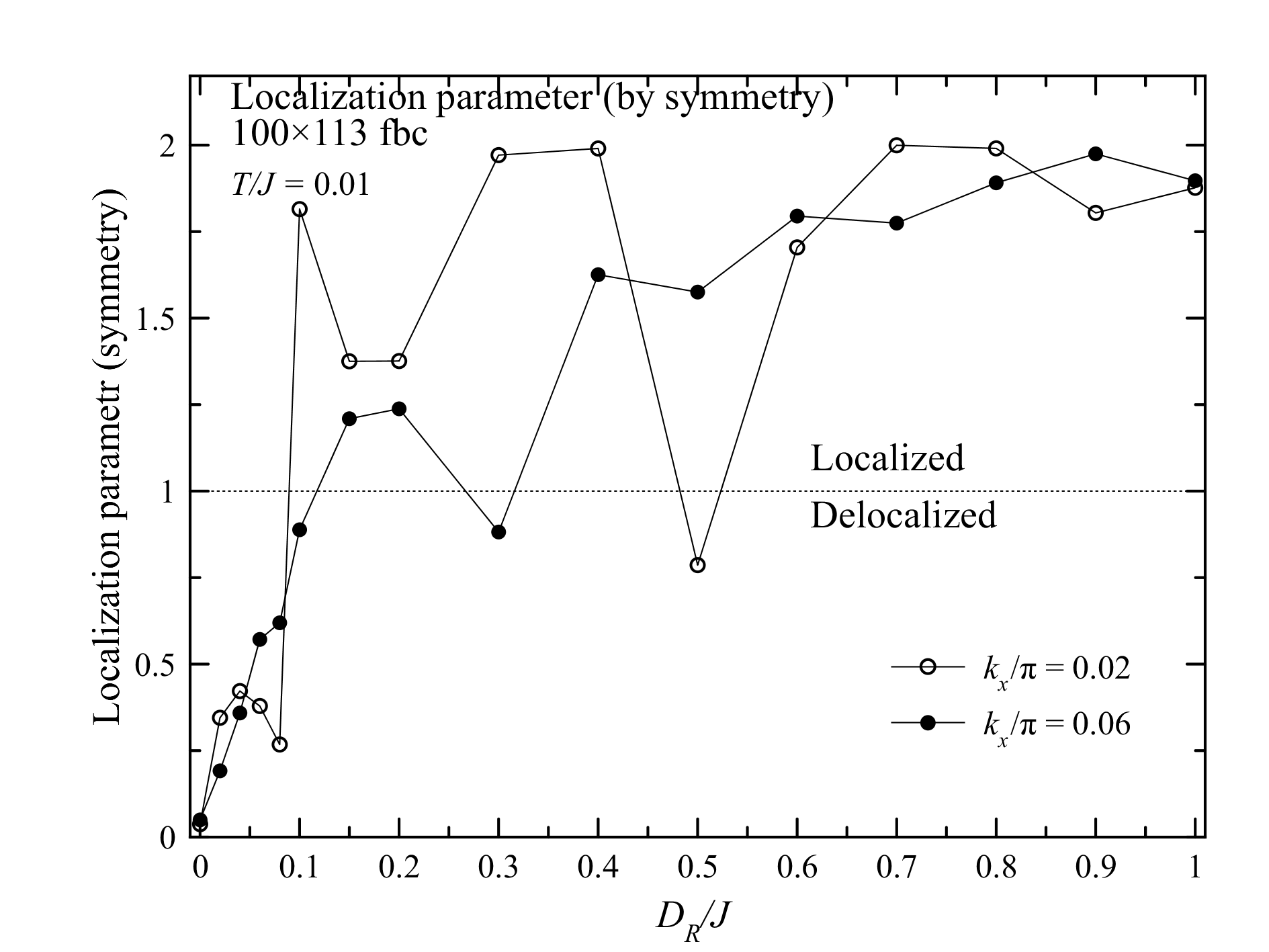}
\par\end{centering}
\caption{Localization parameter $\zeta$ vs $D_{R}/J$ for a system of the
size $100\times113$ and the frequencies corresponding to the pure
$k_{x}/\pi=0.02$ and 0.06 modes.}
\label{Fig-Locsymm vs DR 100x113}
\end{figure}

\section{Numerical results for the localized modes}

Numerical calculations were performed with Wolfram Mathematica using
compilation and parallelization. The system sizes studied ranged from
$10^{4}$ to $10^{6}$ spins. Computations for different values of
$D_{R}$ were done in parallel. We used three different computers
with 8, 8, and 16 cores.

\subsection{Visualization of excitation modes}

Fig. \ref{Fig-Modes_pure} shows spatial profiles of spin fluctuations
$P_{i}(\omega)$, Eq. (\ref{Pi_Def}), in the pure $100\times113$
system at $H=0$ and $T/J=0.01$ at two different frequencies matching
those of spin-wave modes, Eq. (\ref{omega_k_Def}). As expected, these
profile are close to the squares of the spin-wave eigenfunctions,
Eq. (\ref{f_eigen_Def}), for the corresponding $k$. In the system
of square shape there is a degeneracy of SW modes with $\mathbf{k}$
along the 10 and 01 directions in the lattice. Thus spin fluctuations
can contain superpositions of the corresponding SW eigenstates with
the weights depending on the state after thermalization. This can
lead to checkered patterns of the spatial profiles of spin fluctuations,
as shown in Fig. \ref{Fig-Modes_pure_superposition}. In particular,
in the upper panel this is the square of the antisymmetric superposition
of the two states in Eq. (\ref{F_eigen_Def}), $F_{k,n_{x}n_{y}}^{2}=\left(f_{n_{x}k}-f_{n_{y}k}\right)^{2}$.
In the lower panel there is apparently a superposition of a larger
number of modes.

Fig. \ref{Fig-Modes_100x100_DR=00003D1} shows spatial profiles of
spin fluctuations $P_{i}(\omega)$, Eq. (\ref{Pi_Def}), in the RA
system with $D_{R}/J=1$ at $H=0$ and $T/J=0.01$ at two different
frequencies matching those of spin-wave modes, Eq. (\ref{omega_k_Def}).
Here, unlike the results for the pure model shown in Fig. \ref{Fig-Modes_pure},
spatial spin profiles are irregular and localized that implies localization
of spin waves in the presence of randomness. This effect was detected
in Ref. \citep{GC-PRB2021} by exciting the 2D RA magned by short
(five periods) pumping by a weak time-dependent magnetic field at
particular frequencies (see Fig. 11 of this paper). The response to
the pumping was localized in space as narrow peaks that oscillated
in time. Here local modes are investigated more accurately by making
a Fourier transform of thermal spin fluctuations over long times.
At such times, the spatial profiles stabilize and only peaks oscillating
with the frequency close to the frequency set in the Fourier transformation
are selected.

In contrast to the standing waves in pure model, the excitation modes
in the RA system are apparently localized. The shape and positions
of the peaks are irregular. Also, the peaks are rarified. This can
be explained by the redistribution of the density of states. In the
pure model, $P_{i}(\omega)$ has an appreciable value if $\omega$
matches one of the quantized spin-wave frequencies $\omega_{\mathbf{k}}$
and a very small value otherwise. In the random model with $D_{R}/J\sim1$,
one can expect that the frequencies of the modes are not quantized
and take any values in a particular range.

\begin{figure}
\begin{centering}
\includegraphics[width=9cm]{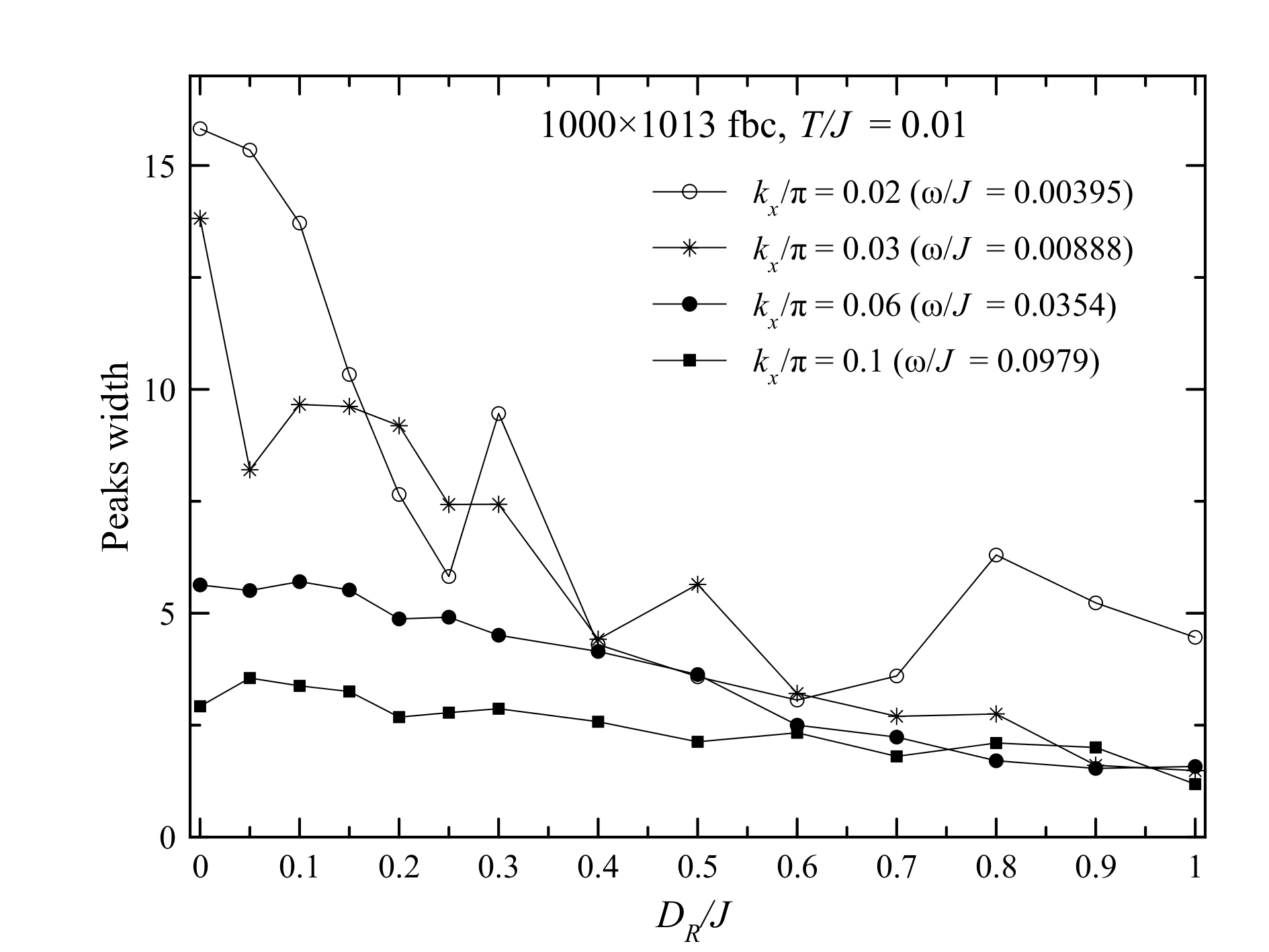}
\par\end{centering}
\begin{centering}
\includegraphics[width=9cm]{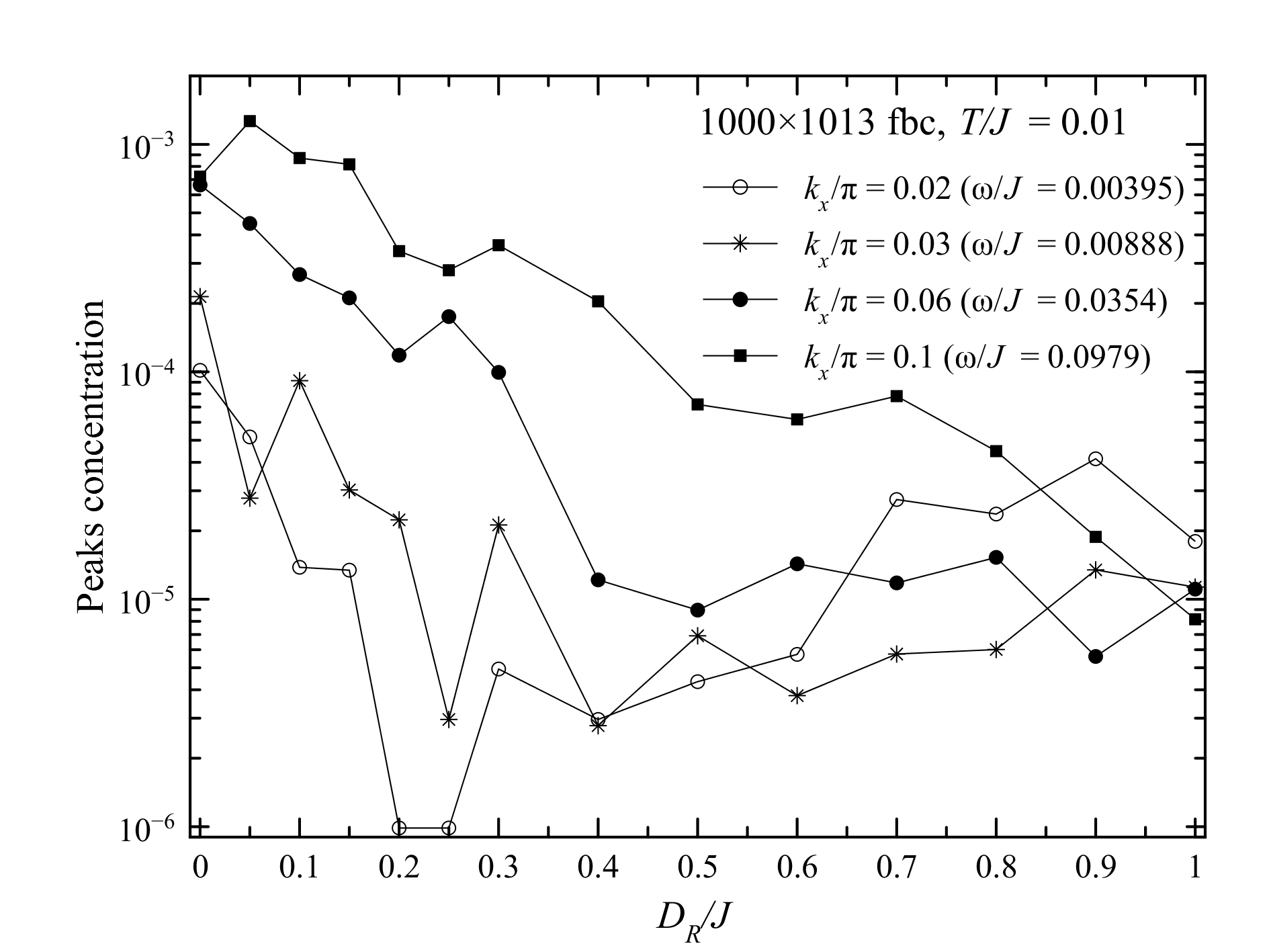}
\par\end{centering}
\caption{Average peaks' widths $\delta$ (upper panel) and peaks' concentration
$N_{p}$ (lower panel) vs the RA strength $D_{R}$ at different frequencies
for the $1000\times1013$ system.}

\label{Fig-Peaks-1000x1013}
\end{figure}
\begin{figure}
\begin{centering}
\includegraphics[width=9cm]{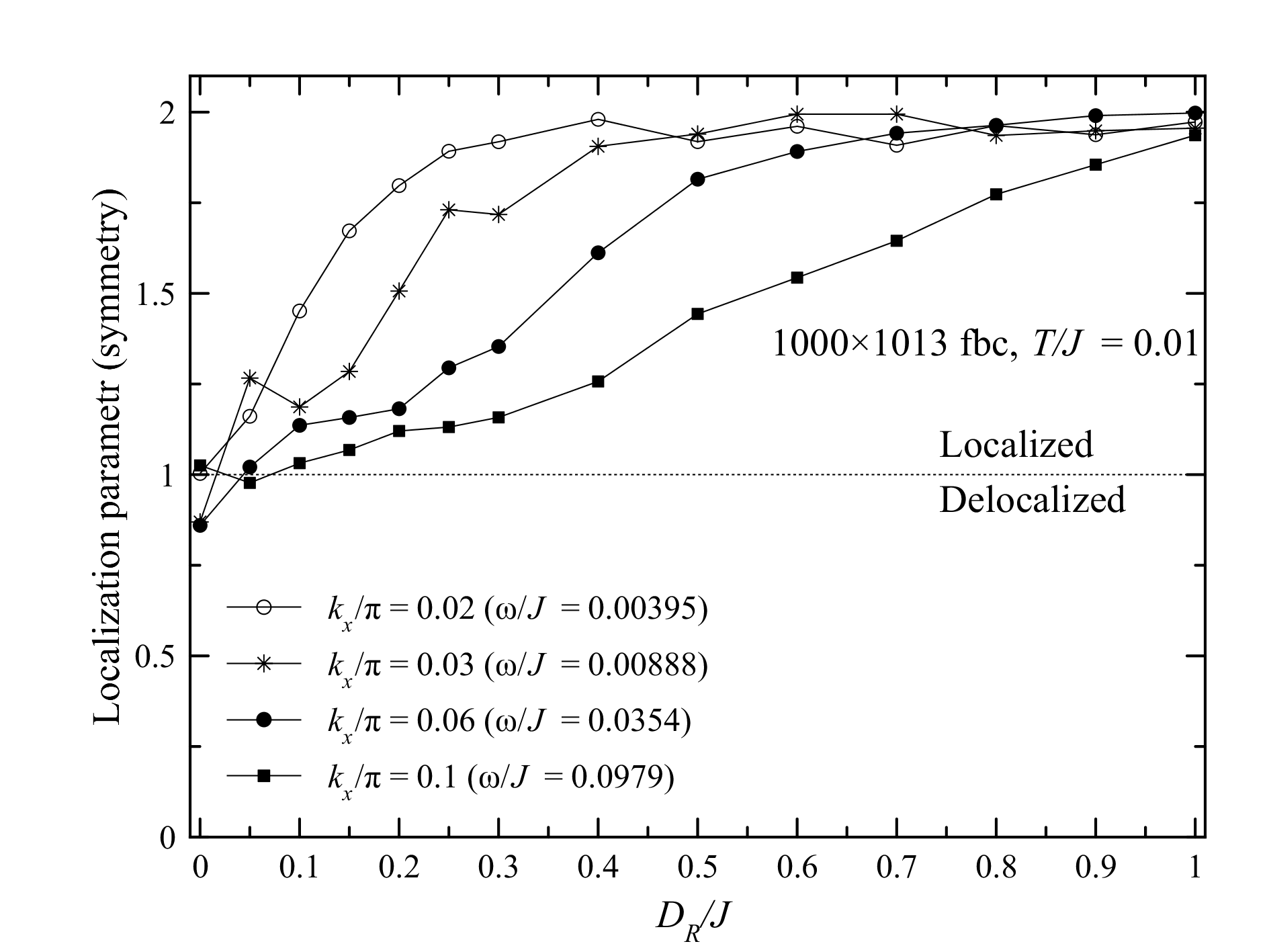}
\par\end{centering}
\caption{Localization parameter $\zeta$ vs $D_{R}/J$ at different frequencies
for a system of the size $1000\times1013$.}

\label{Fig-locsymm-1000x1013}
\end{figure}

\subsection{Investigation of the localized modes parameters}

The results for the average peak width vs the integration time up
to $t=70000$ in Eq. (\ref{s_Fourier}) in periods of the frequency
corresponding to the $k_{x}/\pi=0.06$ mode in the pure system, obtained
by the two methods, are shown in Fig. \ref{Fig-PeakWidth_vs_nT0}.
The curves in the upper panel for the pure system show gradual selection
of the standing wave along the $x$-axis of the type shown in Fig.
\ref{Fig-Modes_pure}. A significant difference between the peak widths
computed by the two methods is due to the washboard profile with no
peaks in this case. In the presence of RA in the lower panel, the
results obtained by the two methods are in a good accordance. The
results do not significantly change with the evolution time and there
are significant fluctuations that do not die out with the time {[}see
comments below Eq. (\ref{Pi_omega_analytical}){]}. 

For larger system sizes, the spin-wave modes becomes densely spaced,
so that there are modes with many different $\mathbf{k}$ having nearly
the same frequency. This makes selection of only one mode, as in Fig.
\ref{Fig-Modes_pure}, impossible. Instead, one obtains a complicated
landscape with many peaks due to the superposition of many different
standing waves.

The results for the average peak width at the end of the evolution
obtained with the first method are shown in Fig. \ref{Fig-Peaks_widths_vs_DR}
One can see that the peaks become narrower for larger $D_{R}$. The
scatter is rather large for this moderate system size. 

The results for symmetry localization parameter $\zeta$ for the system
of $100\times113$ spins with fbc obtained by the solution of the
equation of motion within the time interval $t_{\max}=50000$, see
Eq. (\ref{s_Fourier}), are shown in Fig. \ref{Fig-Locsymm vs DR 100x113}.
The data scatter is large for this moderate system size. The results
below were obtained from the same computation of the evolution of
the thermal spin state.

Numerical results for our largest system of one million spins, $1000\times1013$
are shown in Figs. \ref{Fig-Peaks-1000x1013} and \ref{Fig-locsymm-1000x1013}.
The first figure shows the dependences of both the average peaks'
width $\delta$ and peaks' concentration $N_{p}/N$ on $D_{R}$ at
different frequencies $\omega$ . Both quantities decrease with $D_{R}$
and weakly depend on $\omega$ in the frequency range under investigation.
These results for the peaks' width are in accord with those for our
small system, $100\times113$, in Fig. \ref{Fig-Peaks_widths_vs_DR}.
The results for the symmetry localization parameter $\zeta$ in Fig.
\ref{Fig-locsymm-1000x1013} differ from those in Fig. \ref{Fig-Locsymm vs DR 100x113}
as $\zeta$ approaches one rather than zero in the pure limit. This
can be explained by the large amount of different modes with frequencies
$\omega_{\mu}$ close to $\omega$ a the large system. Superposition
of these modes with random phases does not possess a clear symmetry
that is needed to render $\zeta=0$. As in the pure limit excitation
modes are delocalized, the peaks of $P_{i}(\omega)$ are broad is
no empty space between them. For this reason, symmetry operations
above result in annihilation and creation of peaks at the same rate
that yields $\zeta=1$. With increasing $D_{R}$, peaks become rarified,
so that they do not overlap with annihilation as the result of symmetry
operations, and the number of peaks doubles, leading to $\zeta=2$.

\subsection{Investigation of the modes' phases}

\begin{figure}
\begin{centering}
\includegraphics[width=9cm]{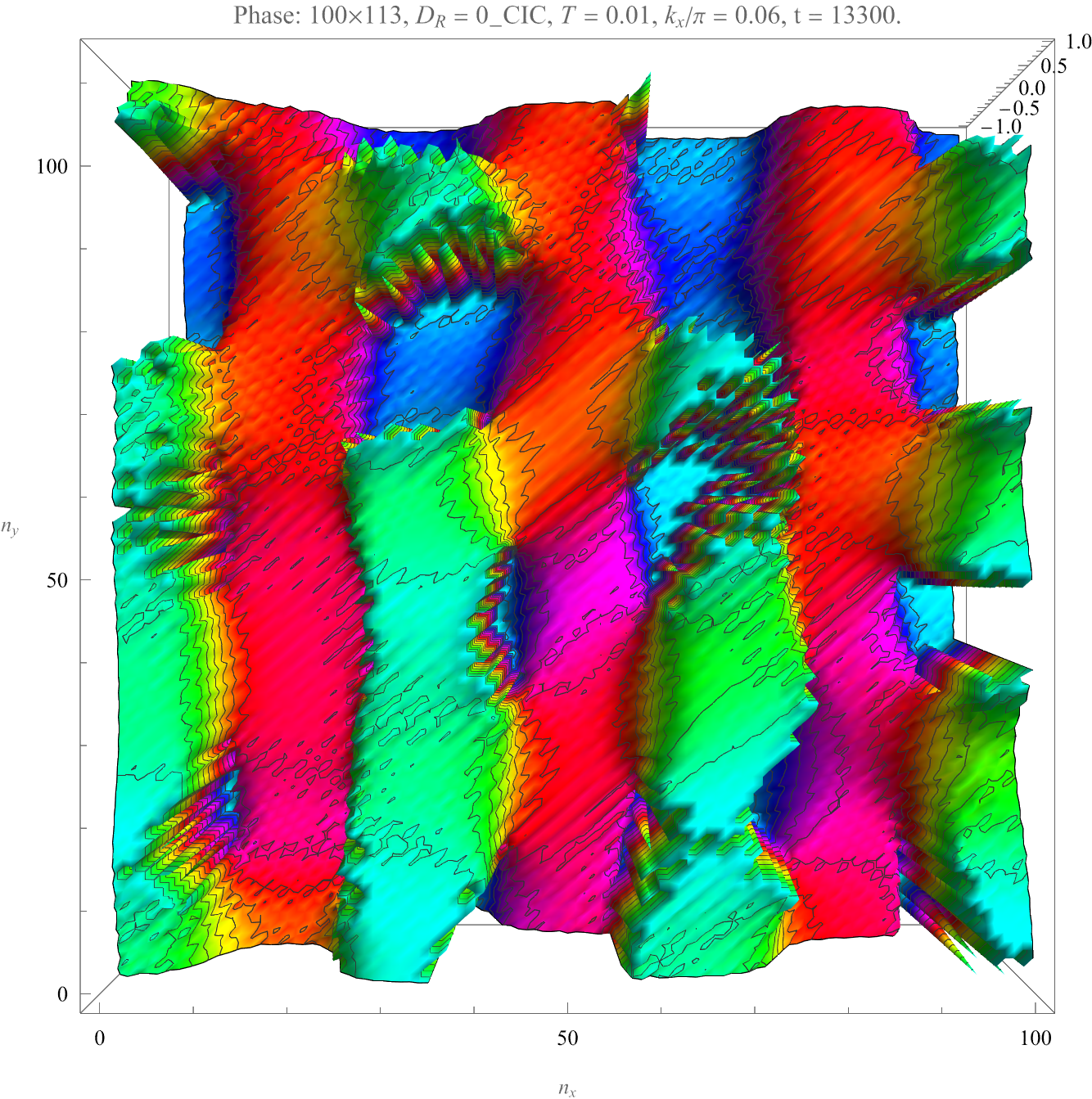}
\par\end{centering}
\caption{A typical phase landscape of a system of standing spin waves in a
pure ($D_{R}=0$) $100\times113$ system color-coded by the Hue function.}

\label{Fig-Phase_pure}
\end{figure}

\begin{figure}
\begin{centering}
\includegraphics[width=9cm]{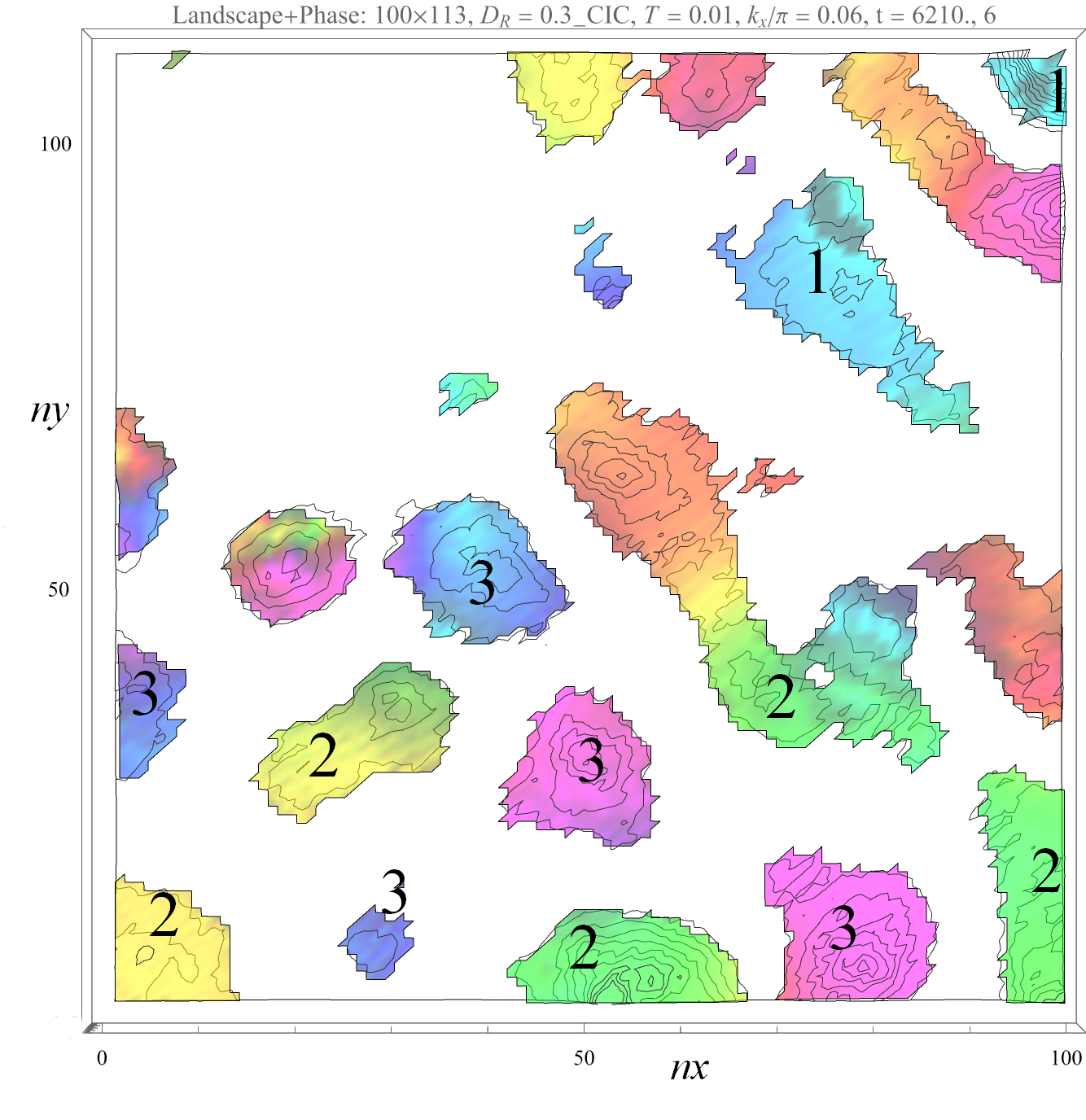}
\par\end{centering}
\begin{centering}
\includegraphics[width=9cm]{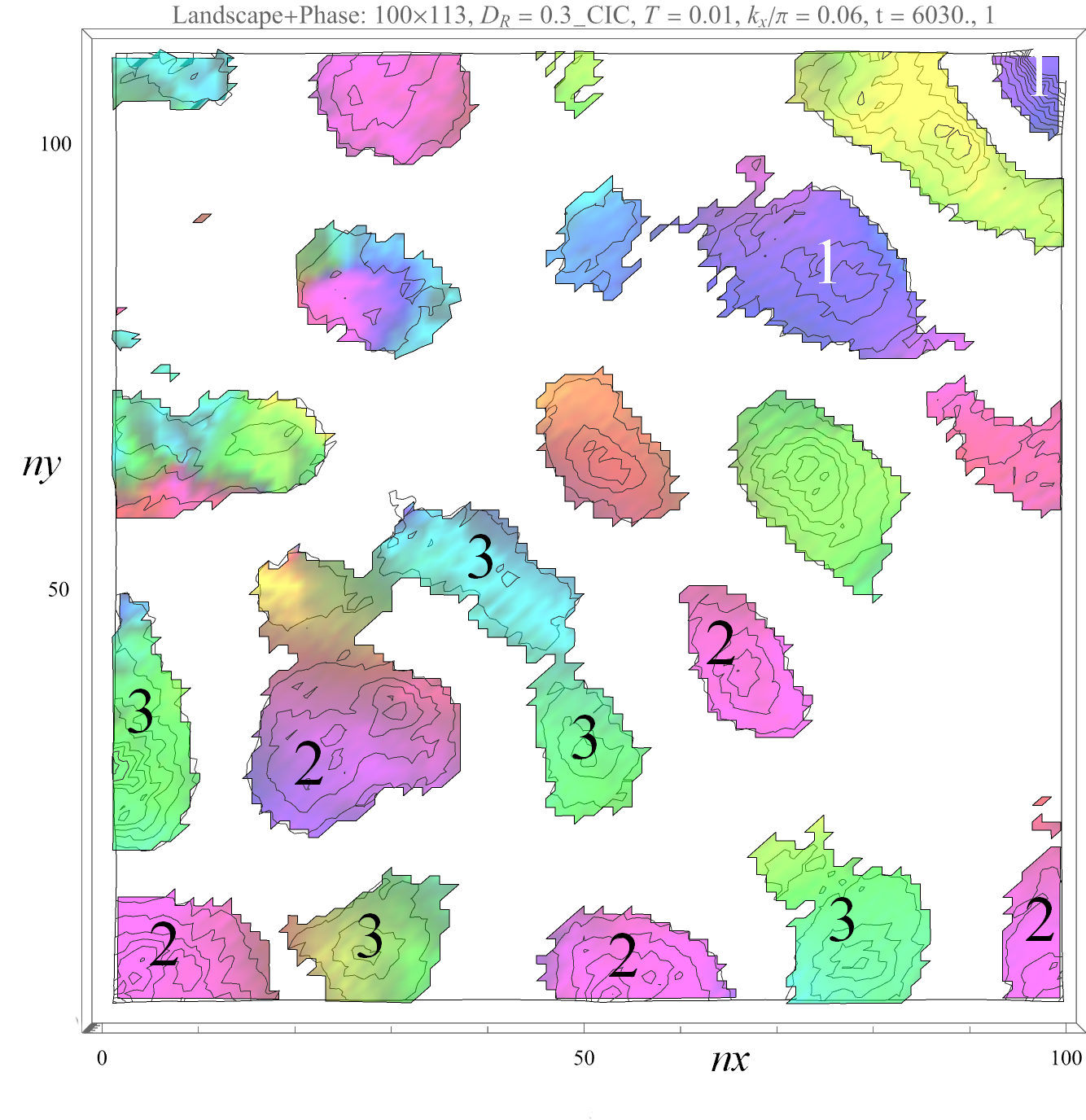}
\par\end{centering}
\caption{Excitation-mode peaks with their phases marked by colors. The upper
and lower panels have the same RA configuration and the same local-energy-minimum
state, however, they have undergone different Monte Carlo procedures
to excite the modes that as the result have different phases. Peaks
labeled by the same color and the same number are in-phase and thus
belong to the same excitation mode.}

\label{Fig-Landscape+phase}
\end{figure}

In this section, we investigate the phase effect in the dynamics of
the excitation modes of a RA magnet by computing and plotting the
phase of the function $p_{i}(\omega)$ defined by Eq. (\ref{pi_Def}).
For the pure model, $D_{R}=0$, there is a system of standing waves
that possesses extended phase correlations of the type shown in Fig.
\ref{Fig-Phase_pure}. Here the phase is represented as a 3D plot
with color coding by by the $\mathrm{Hue\left[\phi/(2\pi)\right]}$
function. The phase is bound to the interval $\left(-\pi,\pi\right)$,
thus the argument of the Hue function is in the $\left(-0/5,0.5\right)$
interval. At the end of this interval, the cyan color is output, and
within this interval, there are all rainbow colors. For the pure system,
the phase is plotted everywhere. For the RA system with localized
modes, the mode intensity $P_{i}(\omega)$ is small outside the peaks,
and in these regions the phase is not well defined. Thus for the RA
system the phase is plotted only in the peaks' regions.

Excitation-mode peaks of $P_{i}(\omega)$ situated at some distance
from each other can be independent local modes or they can be parts
of the same mode. The distinguishing between these cases is subtle
and depends on the coupling between the close mode peaks. The question
of the latter can be investigated with the help of Eq. (\ref{pi_result})
by computing the phase of $p_{i}(\omega)$. The numerical experiment
is the following. For one realization of the random anisotropy, one
finds a local energy minimum. Then one performs parallelized Monte
Carlo thermalization cycles at a very low temperature (here $T/J=0.01)$
for a number of replicas of the system (for instance, equal to the
number of processor cores). Each of these computations leads to different
thermal states with different degrees of excitation of each mode and
their different phases. After that, parallelized computation of the
system's dynamical evolution is run and $p_{i}(\omega)$ are computed
for each replica. If the phase relation between different mode peaks
is the same for different replicas, these peaks belong to the same
mode, otherwise, they are independent local modes. 

An example of the output of this procedure is shown in Fig. \ref{Fig-Landscape+phase}.
In the upper and lower panels, for the $100\times113$ system with
$D_{R}/J=0.3$ the peaks shown by geodesic lines are combined with
the phase at some moment of time, color-coded by the Hue function
as explained above. The two panels are picked out of eight ones resulting
from the parallelized computation. (All eight panels can be seen in
Supplemental Materials). One can see peaks having the same phases
having the same or nearly the same color and labeled by the same number:
1, 2, or 3. For the two different panels, the phases are different,
as it should be given the random nature of the excitation of modes
by Monte Carlo. But the phases are the same within the groups. This
one can conclude that the two peaks labeled by 1 belong the the same
mode and, most likely, the peak between them also belongs to the same
mode. Apart of this, peaks labeled by 2 and by 3 must belong to the
same excitation mode. Likewise, as groups of peaks 2 and 3 are interseeded,
they are likely parts of the same extended mode. To conclude, for
$D_{R}/J=0.3$ there is a system of standing waves distorted by RA
rather than independent localized modes. 

To the contrast, for $D_{R}/J=1$ peaks are narrow and rarified, see
lower panel of Fig. \ref{Fig-Modes_100x100_DR=00003D1}, so that the
interaction between them should be negligible and each peak should
correspond to an individual local mode.

\section{Localization length via edge pumping}

\begin{figure}
\begin{centering}
\includegraphics[width=9cm]{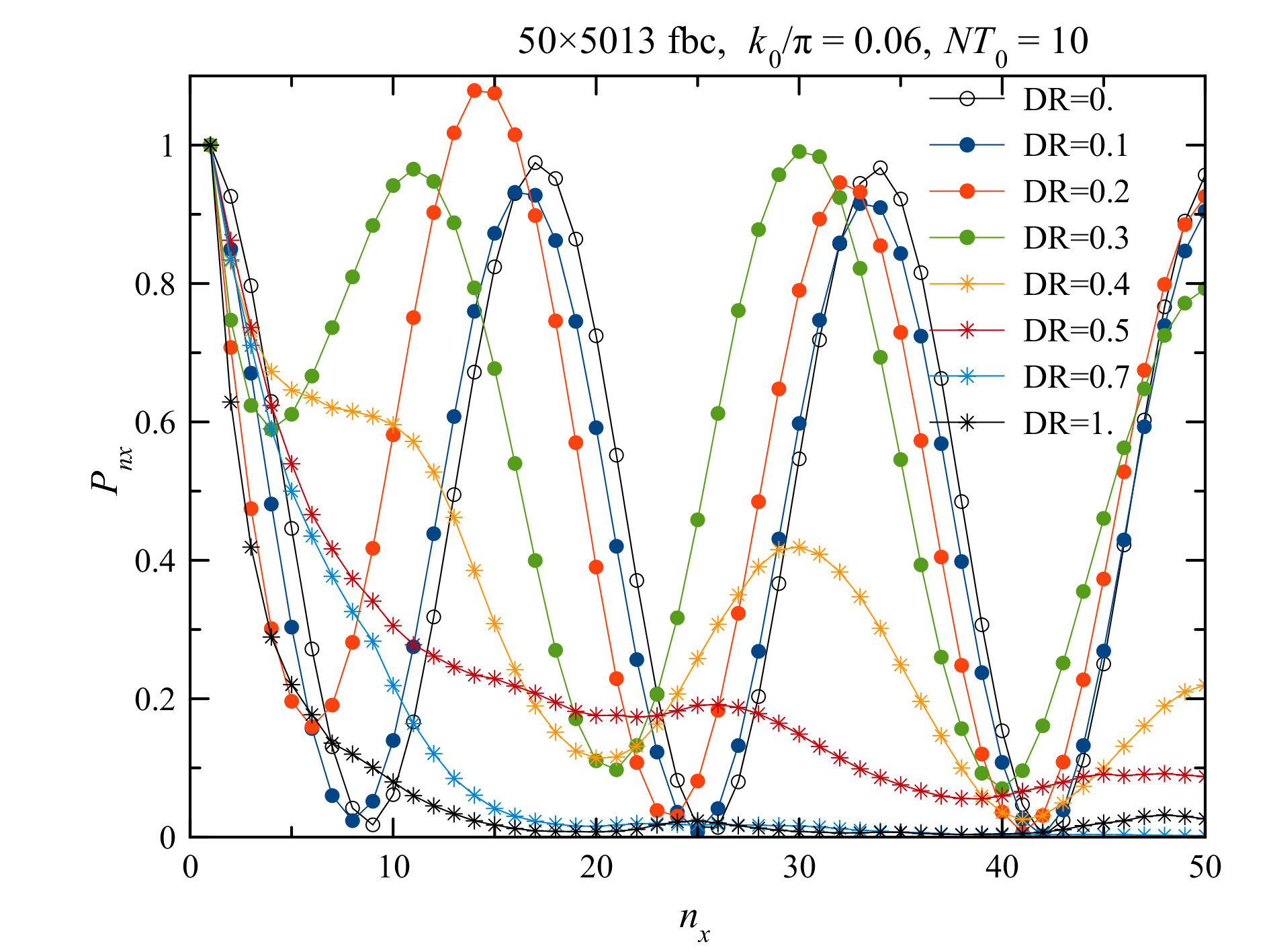}
\par\end{centering}
\caption{The edge-pumping excitation profiles, normalized by their value at
the left boundary, for the $50\times5013$ RA model at the frequency
corresponding to the pure-system standing wave with $k_{0}/\pi=0.06$.
The localization effect becomes strong for $D_{R}/J=0.4$.}

\label{Fig-Edge_pumped_states}
\end{figure}
\begin{figure}
\begin{centering}
\includegraphics[width=9cm]{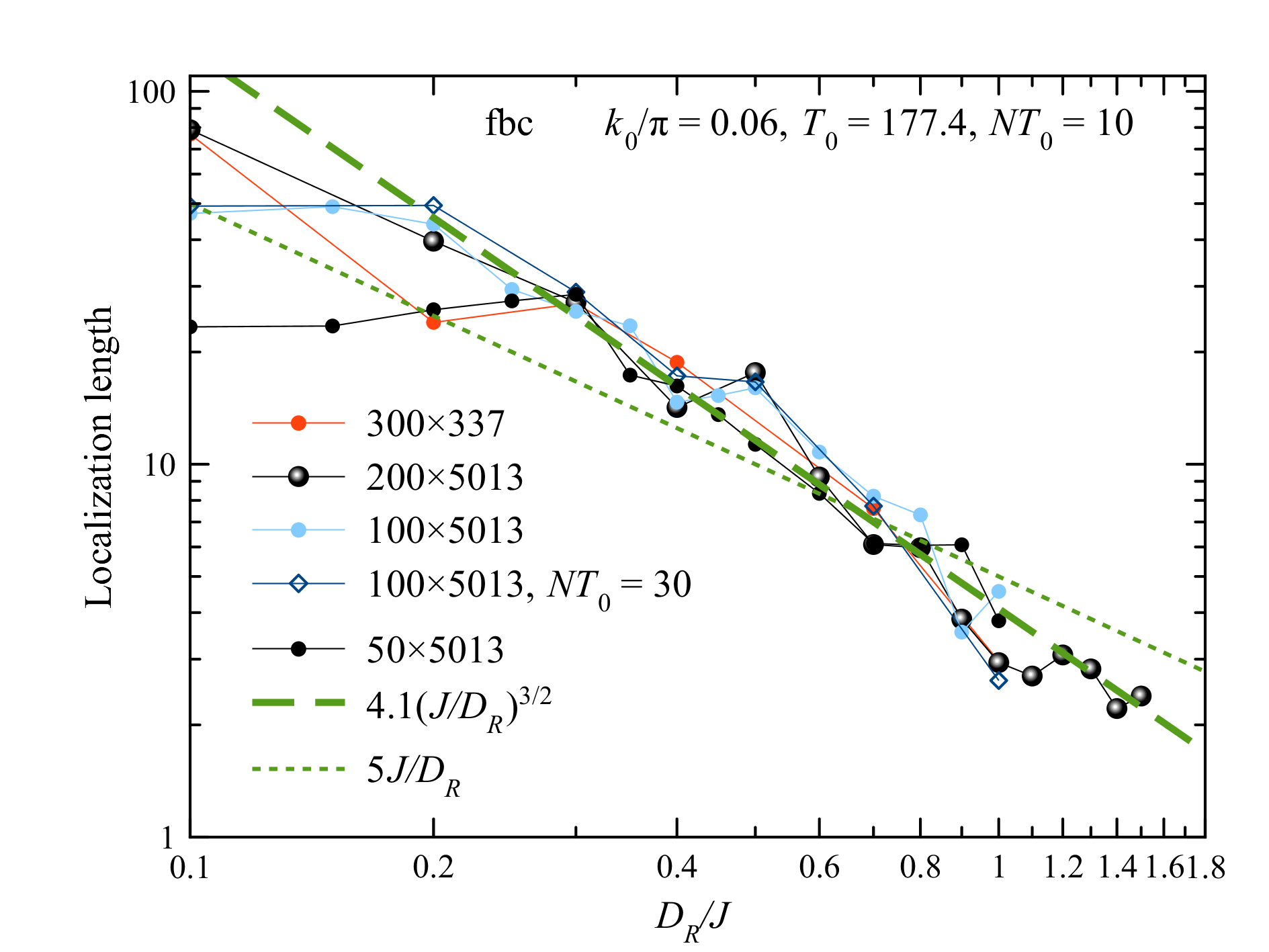}
\par\end{centering}
\caption{Localization length $\xi_{loc}$ vs $D_{R}/J$ for systems of different
sizes at the frequency corresponding to $k_{0}/\pi=0.06$.}

\label{Fig-LocalizationLength}
\end{figure}
\begin{figure}
\begin{centering}
\includegraphics[width=9cm]{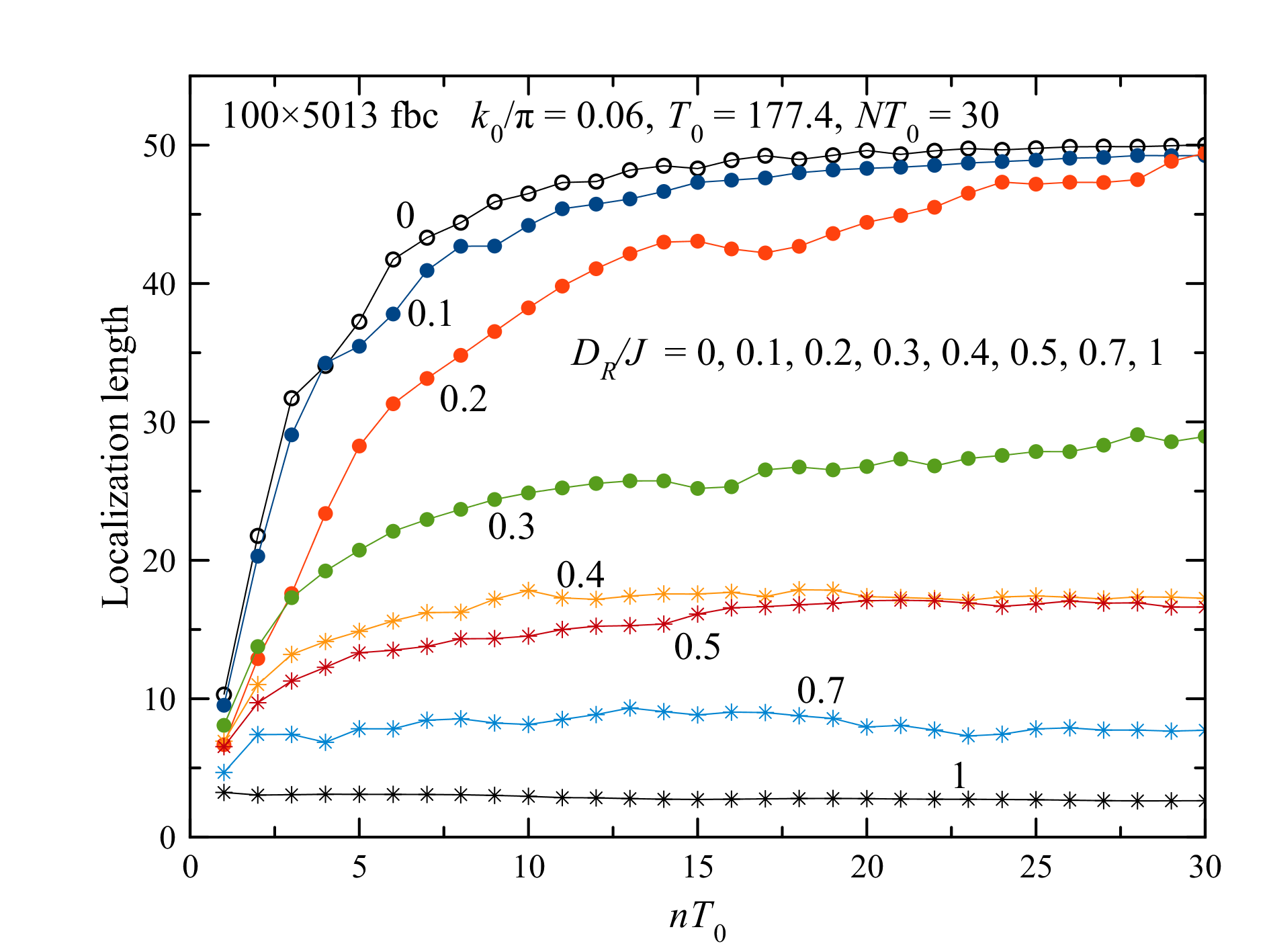}
\par\end{centering}
\caption{The dependence of the computed localization length $\xi_{loc}$ on
the number of periods of pumping $n_{T_{0}}$for the $100\times5013$
system.}

\label{Fig-LocalizationLength_vs_DR_vs_nT0}
\end{figure}

The study of the thermal dynamics of the RA magnet above includes
a characterization of the excitation modes that allows to judge if
they are extended or localized. However, this characterization does
not prove that the modes are indeed localized. Many peaks resonating
at the same frequency may be different localized modes or they can
be the same mode with several peaks. The latter happens if local modes
become hybridized by the interaction or, ar it can be put, there is
tunneling between degenerate or quasi-degenerate local modes.

The localization can be tested, for instance, by the edge-pumping
method. A sinusoidal magnetic field at a frequency $\omega_{0}$ matching
that of one of standing waves is applied to the spins at one of the
boundaries and it is monitored how the excitation is penetrating into
the body of the system. For the pure system, a stading wave of the
matching frequency is excited. In the case of localized modes in the
strong-RA model, the wave is not penetrating into the body of the
system and excitation is limited to the region near the pumped edge.
This experiment allows to estimate the localization length. If the
left boundary, $n_{x}=1$, is pumped, one can define the excitation
profile
\begin{equation}
P_{n_{x}}(\omega)=\delta\bar{\mathbf{s}}_{n_{x}}(\omega)\cdot\delta\mathbf{\bar{s}}_{n_{x}}^{*}(\omega),\label{Edge_pumping_excitation_profile}
\end{equation}
where $\delta\bar{\mathbf{s}}_{n_{x}}(\omega)$ is the average over
the transverse direction:
\begin{equation}
\delta\mathbf{\bar{s}}_{n_{x}}(\omega)\equiv\frac{1}{N_{y}}\sum_{n_{y}=1}^{N_{y}}\delta\mathbf{\tilde{s}}_{n_{x},n_{y}}(\omega),
\end{equation}
whereas $\delta\mathbf{\tilde{s}}_{n_{x},n_{y}}(\omega)\equiv\delta\mathbf{\tilde{s}}_{i}(\omega)$
is given by Eq. (\ref{s_Fourier}). The localization length, in lattice
units, can be estimated as
\begin{equation}
\xi_{loc}=\frac{1}{P_{1}(\omega)}\sum_{n_{x}=1}^{N_{x}}P_{n_{x}}(\omega).\label{xi_loc_def}
\end{equation}
For the pure system it is close to $N_{x}/2$ but decreases with increasing
$D_{R}$. It turns out that regions inside the system may respond
stronger to pumping than the left boundary because these local modes
are at resonance with the pumping whereas the left-boundary region
is not. However, choosing a very large $N_{y}$ one can average these
effects out. Still, even at large $N_{y}$, the response $\delta\mathbf{\tilde{s}}_{n_{x}}(\omega)$
for small $D_{R}/J$ can be larger somewhere away from the boundary.
To deal with this, it is better to replace $P_{1}(\omega)\Rightarrow\max\left(P_{n_{x}}(\omega)\right)$
in the definition of $\xi_{loc}$. 

As an illustration, the edge-pumping excitation profiles for the $50\times5013$
RA model at the frequency corresponding to the pure-system standing
wave with $k_{0}/\pi=0.06$ are shown in Fig. \ref{Fig-Edge_pumped_states}.
Here, for $D_{R}/J\leq0.3$ the wave is weakly damped and penetrates
into the whole system. Notice that for $D_{R}/J=0.2$ the maximum
at $n_{x}\simeq15$ is higher than the value at $n_{x}=1$, thus it
make sense to use this value for the normalization instead of $P_{1}(\omega)$
in Eq. (\ref{xi_loc_def}). 

The dependence of the localization length on $D_{R}$ is shown in
Fig. \ref{Fig-LocalizationLength} for systems of different sizes
and the same frequency corresponding to $k_{0}/\pi=0.06$. Whereas
for small $D_{R}$ the result depends on the system length $N_{x}$
as expected, for stronger RA all point merge to a size-independent
curve indicating the localization. As, especially for large $N_{x}$,
it takes some time to form a standing wave, pumping should last for
a number of periods $N_{T_{0}}$of the pumping frequency. In Fig.
\ref{Fig-LocalizationLength} we used $N_{T_{0}}=10$. The fitted
dependence of the localization length is $\xi_{loc}=4.1\left(J/D_{R}\right)^{3/2}$.
This differs from the dependence $R_{f}\propto J/D_{R}$ that follows
from Eq. (\ref{Rf_IM}) in two dimensions.

The dependence of the computed localization length $\xi_{loc}$ on
the number of periods of pumping $n_{T_{0}}$up to $n_{T_{0}}=30$
is shown in Fig. \ref{Fig-LocalizationLength_vs_DR_vs_nT0}. One can
see that the values of $\xi_{loc}$ asymptotically saturate but for
smaller $D_{R}$ it takes a longer time. For $D_{R}/J=1,$ the asymptotic
valus is reached immediately as the excitation mode is strongly localized.
One result for $N_{T_{0}}=30$ is shown in Fig. \ref{Fig-LocalizationLength}
but the difference with the $N_{T_{0}}=10$ result is nonessential.

The numerical results show that the localization length $\xi_{loc}$
is independent of the pumping frequency within the numerical scatter.
It is close to the peak width computed from the dynamics of the thermally
excited state, see Fig. \ref{Fig-Peaks_widths_vs_DR}. The latter
also does not depend on the frequency. Thus we conclude that there
is a sufficient evidence of the localization of exiotation modes in
the RA model.

There has to be a relation between the localization of spin waves
and the spin-wave damping. The time decrement $\Gamma$ (relaxation
rate) should be related to the spatial decrement $\kappa$ by the
same relation as frequency and wave vector, that is, $\Gamma_{k}=v_{k}\kappa$,
where $v$ is the spin-wave group velocity $v_{k}=d\omega_{k}/dk$.
From Eq. (\ref{omega_k_Def}) in the long-wavelength limit one obtains
$\hbar\omega_{\mathbf{k}}=4J\left(ak\right)^{2}$ and thus $v_{k}=8\left(J/\hbar\right)a^{2}k$.
With $\kappa=1/\xi_{loc}$ one obtains $\Gamma_{k}=8\left(J/\hbar\right)a^{2}k/\xi_{loc}$.
If $\xi_{loc}$ does not depend on $k$, then $\Gamma_{k}\propto k$. 

\section{Conclusion}

\label{discussion}

We have studied localized spin excitations in an RA magnet. Our main
results can be summarized as follows. Standing spin waves in a finite-size
sample evolve into localized modes on increasing the strength of the
RA. We visualized this process by taking snapshots of spin oscillations
in a 2D system. The study of phases of localized excitations shows
that they generally consist of a group of well-separated peaks oscillating
coherently. The average width of the peak, which we identify with
the localization length, decreases on increasing the strength of the
RA, scaling roughly as $(J/D_{R})^{3/2}$.

Notice that in the absence of topological defects, the ferromagnetic
correlation length in a 2D RA magnet scales as $J/D_{R}$. However,
in a 2D Heisenberg model with three-component spins, the effect of
topological defects (skyrmions in particular) must be significant
\citep{PGC-PRL2014}. It must lead to the decay of ferromagnetic correlations
on a shorter scale that is consistent with the $(J/D_{R})^{3/2}$
dependence on the RA strength and the suggestion \citep{GC-PRB2021}
that localized excitations in the RA magnets are hosted by the Imry-Ma
domains.

Broad distribution on the effective magnetic anisotropy for Imry-Ma
domains is responsible for the broad distribution of spin-precession
frequencies in the RA magnets. Combined with the high density of the
Imry-Ma domains, it provides strong broadband absorption of microwave
power. The width of the band rapidly increases on increasing the strength
of the RA. Our studies of the excitation of localized modes by the
ac magnetic field establish a connection between localization and
power absorption.

Based upon dissipation-fluctuation theorem, one can show that the
integral power absorption in an RA magnet, $\int_{-\infty}^{\infty}d\omega P(\omega)$,
is proportional to $D_{R}^{2}$. Combined with the dependence of the
localization length on $D_{R}$ shown in Figs. \ref{Fig-Peaks_widths_vs_DR},
\ref{Fig-Peaks-1000x1013}, and \ref{Fig-LocalizationLength}, it
indicates that the integral power scales inversely with the localization
length. Earlier we demonstrated \citep{GC-PRB2021} that the height
of the peak in $P(\omega)$ has a weak dependence on the RA strength.
This makes the integral power a measure of the absorption bandwidth.
The results obtained in this paper show that the broadband absorption
of microwaves by random magnets is intimately related to the localization
of spin excitations: The stronger the localization the greater the
bandwidth. 

It would be interesting to visualize the localization of spin oscillations
in an amorphous or sintered ferromagnetic film in a real experiment
that would generate images similar to the ones shown in Fig. \ref{Fig-Modes_100x100_DR=00003D1}.
This could be done by, e.g, covering the film with a fluorescent layer
that reacts to the inhomogeneous absorption of microwave power by
the magnetic layer.

\section*{Acknowledgements}

This work has been supported by Grant No. FA9550-20-1-0299 funded
by the Air Force Office of Scientific Research.\\

\end{document}